\documentclass[12pt]{article}

\pdfoutput=1

\usepackage{amsmath,amssymb,amsfonts,amscd,mathrsfs}
\usepackage{xcolor}
\definecolor{darkblue}{rgb}{0.1,0.1,.7}
\usepackage[colorlinks, linkcolor=darkblue, citecolor=darkblue, urlcolor=darkblue, linktocpage]{hyperref} 
\usepackage[square, comma, compress,numbers]{natbib}
\usepackage[]{graphicx}

\usepackage{geometry}
\usepackage[margin=10pt,font=small,labelfont=bf]{caption}
\usepackage{ifthen}
\usepackage{tikz}
\usepackage{subcaption}
\usepackage{booktabs,multirow}
\usepackage{hhline}
\usepackage{tablefootnote}

\usepackage{ragged2e}
\usepackage{array}
\newcolumntype{L}[1]{>{\raggedright\let\newline\\\arraybackslash\hspace{0pt}}m{#1}}
\newcolumntype{C}[1]{>{\centering\let\newline\\\arraybackslash\hspace{0pt}}m{#1}}
\newcolumntype{R}[1]{>{\raggedleft\let\newline\\\arraybackslash\hspace{0pt}}m{#1}}

\usepackage{booktabs}

\usepackage{titlesec}
\titleformat*{\section}{\large\bfseries}
\titleformat*{\subsection}{\normalsize\bfseries}
\titleformat*{\subsubsection}{\normalsize\it}
\titleformat*{\paragraph}{\normalsize\bfseries}
\titleformat*{\subparagraph}{\normalsize\bfseries}

\def\bit{\begin{itemize}}
\def\eit{\end{itemize}}
\def\baa{\begin{array}}
\def\eaa{\end{array}}

\newcommand{\reef}[1]{(\ref{#1})}

\def\eps{\epsilon}

\newcommand{\beq}{\begin{equation}} 
\newcommand{\eeq}{\end{equation}}

\def\geq{\geqslant}
\def\leq{\leqslant}
\newcommand{\diffop}[2]{\ifthenelse{\equal{#2}{1}}{\frac{\mrm{d}}{\mrm{d} #1}}{\frac{\mrm{d}^#2}{\mrm{d} #1^#2}}}

\newcommand{\mrm}[1]{{\mathrm #1}}

\usepackage[normalem]{ulem}

\newcommand{\be}{\begin{equation}}
\newcommand{\ee}{\end{equation}}
\newcommand{\bea}{\begin{eqnarray}}
\newcommand{\eea}{\end{eqnarray}}

 \def\om{\omega}
  \def\th{\theta}

\usepackage{accents}
\newlength{\dhatheight}

\numberwithin{equation}{section}
\usepackage[tocgraduated]{tocstyle}

\begin{document}

\vspace*{-.6in} \thispagestyle{empty}
\begin{flushright}
 SISSA  35/2017/FISI
\end{flushright}
\vspace{1cm} {\Large
\begin{center}
{\bf 
{
 Novel measurements of anomalous\\[.2cm]triple gauge couplings for the LHC}
}
\end{center}}
\vspace{1cm}

\begin{center}
{ \bf A. Azatov$^{a,b}$, J. Elias-Mir\'o$^{a}$,  Y. Reyimuaji$^{a,b}$,  E. Venturini$^{a}$} 

{
$^{a}$ SISSA/ISAS and INFN, I-34136 Trieste, Italy\\[.2cm]
$^{b}$ Abdus Salam International Centre for Theoretical Physics, I-34151 Trieste, Italy 
}
\vspace{1cm}
\end{center}

\vspace{4mm}

\begin{abstract}
Finding better ways to prove the Standard Model Effective Field Theory is a very important direction of research. 
This paper focuses on measurements of Electroweak triple gauge couplings, paying special attention on the regime of validity of the Effective Field Theory (EFT). In this regard, one of our goals is to find measurements leading to a large increase of the interference  between the SM amplitude and the contribution of irrelevant operators in the EFT.  We propose two such distributions that will lead to a better accuracy.  
Improvements compared to the traditional methods as well as LHC high luminosity prospects are discussed.

 \end{abstract}
\vspace{.2in}
\vspace{.3in}
\hspace{0.7cm}  

\newpage

\setcounter{tocdepth}{2}

{
\tableofcontents
}

\section{Introduction}

The Standard Model (SM) of particle physics is our best model describing the innermost layer of matter. It has been verified in uncountable experiments spanning a wide range of energies. The Higgs discovery \cite{Aad:2012tfa,Chatrchyan:2012xdj} was the icing of the cake of more than forty decades of experiments confirming every testable prediction of the SM. Now, the most important goal of the LHC is the quest for new physics, either  in the form of  deviations from the SM predictions or as new degrees of freedom in direct searches. 

ATLAS and CMS  have performed many dedicated searches of beyond the Standard Model (BSM) theories \cite{talksEPS}. All such investigations have led to null results. 
Before the run of these experiments it was widely acknowledged that the confirmation of the SM and nothing more is a logical possibility. At the same time though there are many theoretically appealing BSM extensions that seem to make sense. Thus, \emph{why nature is not making use of them?} is a very pressing question that should have an answer. 
In order to make progress towards answering this question we can envision two possible strategies: more clever model building  -- which may require a paradigm change with respect to conventional views; or to understand in detail the real pressure that the LHC is imposing on the BSMs. This work deals with a particular example in the second  direction. 

The experimental results suggest that there is at least a moderate mass gap between the electroweak scale $m_W$ and the new physics scale $\Lambda$. 
Given this situation it is very convenient to parametrize possible deviations from the SM in an EFT approach. 
This consists in viewing the SM as the leading interactions of an effective Lagrangian 
and incorporate BSM deviations in a perturbative expansion in powers of SM  fields or derivatives $D_\mu$ over the proper power of $\Lambda$,
\be
{\cal L}_\text{eff} =
{\cal L}_\text{SM} +{\cal L}_6+ \cdots \, , \label{leff}
\ee
where ellipses denote terms of order $1/\Lambda^3$ and higher. 
Given the uncertainty of the current situation we will take a skeptical point of view on the particular UV physics leading to \reef{leff} and thus only assume the SM gauge symmetries.
Then, up to the dimension five Weinberg operator $\sim \Psi_L \Psi_L H H$,  the leading deviation from the SM consists in  operators of dimension six,
\be
{\cal L}_6 = \sum_i  \frac{c_iO_i}{\Lambda^2} \, . \label{dim6}
\ee
The dimensionless coefficients $c_i$ are the Wilson coefficients, which we assume to be perturbative but otherwise arbitrary.
The operators appearing in  \reef{dim6} were exhaustively listed in \cite{Grzadkowski:2010es}, see also \cite{Buchmuller:1985jz}.
The  advent of the LHC, especially after the Higgs discovery, has triggered an abundant number  of works on interpreting the LHC searches as limits on effective field theory deformations of the SM. 
It is very interesting to find better ways to measure the SM EFT. This is in fact the purpose of this work, which focuses on  diboson production $WZ/WW$ at the LHC  and how it can be used to constrain the  deformations from the SM due to the triple gauge couplings (TGCs) in ${\cal L}_6$.

In the SM the TGC are fixed by the Lorentz symmetry and given by
\be
 i g \,  W^{+\, \mu\nu } W_\mu^- W_\nu^3+  ig \, W^{3\, \mu\nu} W_\mu^+ W_\nu^- \, ,    \label{smtgc}
\ee
where $W^3_\nu= c_\th\,   Z_{\nu} + s_\th\, A_{\nu}$ is  a linear combination of the Z and photon vector boson, and $\th$ is the Weinberg angle. The interaction in  \reef{smtgc} is written in the unitary gauge, so that the vector boson fields describe both longitudinal and transverse polarizations. 
There are only two types of CP-even anomalous triple gauge couplings (aTGCs) deviating  from \reef{smtgc}. The first one consists in deforming \reef{smtgc} away from the SM point
\be
{\cal L}^{1st}_{aTGC} =  i g \,   c_\th\,   \delta g_{1,Z}\, Z_{\nu}  W^{+\, \mu\nu } W_\mu^- +h.c. + ig \, (c_\th \,   \delta \kappa_Z  \, Z^{\mu\nu} + s_\th\,   \delta \kappa_\gamma \, A^{\mu\nu} ) W_\mu^+ W_\nu^- \, . \label{type1}
\ee
Modifications of the  coupling  $W^{+\, \mu\nu } W_\mu^-  A_\mu$  is forbidden by gauge invariance and the relation   $\delta \kappa_Z= \delta g_{1,Z}- \tan^2 \th \delta \kappa_\gamma$ is satisfied if only dimension six operators are considered.
The other type of deformations are obtained by  adding extra derivatives on \reef{smtgc}. This translates into higher powers of momentum in the amplitudes. In an expansion in powers of momentum, the leading such deformation is 
\be
{\cal L}^{2nd}_{aTGC} =    \lambda_Z  \,  \frac{i g }{m_W^2}\, W^{+\, \mu_2 }_{\mu_1} W^{-\, \mu_3 }_{\mu_2}W^{3\, \mu_1}_{\mu_3}   \, . \label{type2}
\ee
The study of the  triplet of deformations $\{\delta g_{1,Z}, \delta \kappa_Z, \lambda_Z \}$ is a classic test of the SM with a long history starting with  the works \cite{Gaemers:1978hg,Hagiwara:1986vm} and continued by \cite{DeRujula:1991ufe,Hagiwara:1993ck,Baur:1994aj,Baur:1994ia}.~\footnote{See for example \cite{Falkowski:2016cxu,Falkowski:2014tna,Campanario:2016jbu,Berthier:2016tkq,Berthier:2015gja,Butter:2016cvz,Dumont:2013wma,Ellis:2014dva,Gavela:2016bzc}
 for recent TGC and EFT analyses.} Famously, the interactions (\ref{type1},\ref{type2}) were bounded with percent level accuracy at the LEP-2 experiment \cite{Schael:2013ita}:  
\be
\lambda_Z\in [-0.059,0.017]\ , \quad  \delta g_{1,Z}\in[-0.054,0.021]\ , \quad \delta \kappa_Z\in[-0.074,0.051]  \ ,
\ee
at 95\% confidence level.  

At the LHC, we would like to exploit the energy growth of (\ref{type1},\ref{type2})  to put stronger bounds on TGCs.  
However it is well known that some of the TGC contributions   have an additional suppression factor at high energy. In particular the leading energy contribution coming from the $\lambda_Z$ TGC does not interfere with SM for any $2\to 2$ process, which makes its measurements  difficult at LHC.  This is consequence of helicity selection rules \cite{Dixon:1993xd,Azatov:2016sqh,Falkowski:2016cxu}, and the result is valid at leading order (LO).
The main point of our paper is to find ways to  overcome  this suppression. We  propose two measurements that  enhance the  interference of the $\lambda_Z$-BSM amplitude  with the SM contribution.   Our ideas will lead to a better measurement of aTGC at LHC.

The paper is organized as follows:
in section~\ref{tgcpropierties} we review the basic physics associated to  the TGC.
 Then, in  section~\ref{sols} we propose two new variables to improve the accuracy. In section~\ref{sec:EFTvalid} we discuss  the challenges of the EFT measurements at the LHC. Then in  sections~\ref{details} and \ref{res} we discuss  our methodology and the results. We conclude and comment on future directions in section~\ref{conclusion}.

\section{Features of TGC mediated amplitudes}
\label{tgcpropierties}

In this section we review simple facts of the diboson production at the LHC. This will allow us to spot measurements that have not been exploited yet  and will lead to better sensitivity on the TGCs.

Di-boson production at the LHC is dominated by the $2\rightarrow 2$ process  $q\bar q \rightarrow WW/WZ$. 
To neatly expose the leading energy growth of this probability amplitudes we use the Goldstone equivalence theorem. Namely, we work with the parametrization
where the transverse gauge-bosons are massless and the  would-be Goldstone bosons in the Higgs doublet describe the longitudinal components of the $W_{\pm}/Z$ gauge bosons. For definiteness of the notation,
\be
{\cal L}_\text{SM} = (D_\mu H)^\dagger D^\mu H +{\cal L}_\text{gauge} +{\cal L}_{\psi}+ V(H) \, , \label{sm}
\ee
where the  $D_\mu H = (\partial_\mu-i g^\prime Y B_\mu - i g T^a W_\mu^a)H$,  with  $T$  the $SU(2)_L$ generators,   $Y=1/2$ and $H^T=  (\sqrt{2}\, G^+, v+ h+i G_0)/\sqrt{2}$.
As usual, the pure gauge sector is given by the field strengths
$
{\cal L}_\text{gauge}=-\frac{1}{4}   W^a_{\mu\nu} W^{a \mu\nu}-\frac{1}{4} B_{\mu\nu} B^{\mu\nu}-\frac{1}{4}  G_{A \mu\nu} G^{A \mu\nu} \, ,
$
the piece ${\cal L}_\psi$ involves the Kinetic terms for the fermions and the Yukawa interactions, and $V(H)=-m^2|H|^2+\lambda |H|^4$. 
 We recall that  Goldstone's equivalence theorem,
\begin{center}
\includegraphics[scale=0.8]{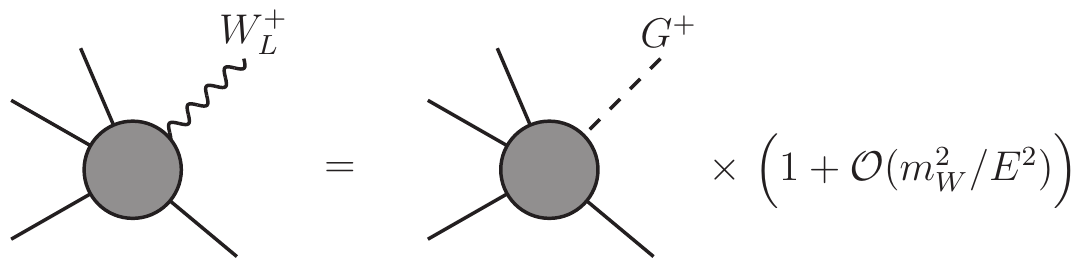} 
\end{center}
states  that to get the leading large energy behavior of the amplitudes with massive gauge bosons in the final state, we can identify in \reef{sm} the transverse and longitudinal components of the physical gauge bosons as 
 \bea
 \{ W_L^+, \, W_T^+ \}  &=& \{ G^+,  \, (W^1-iW^2)/\sqrt{2} \} \ , \label{Wlt} \\ 
 \{ Z_L,  \,  Z_T�\}  &=& \{ G_0/\sqrt{2}  ,\,   \cos \th_w W_3- \sin \th_w \,  B \} \,  , \label{Zlt}
 \eea
 where $\cos\th = g/\sqrt{g^{\prime 2}+ g^2}$ is the cosine of the Weinberg angle. With this basic result in mind, we proceed to discuss the energy growth of diboson production. 
  
\subsection{Energy growth}  With the parametrization in \reef{sm} and the identifications in (\ref{Wlt},\ref{Zlt}), the  SM triple gauge couplings arise from
\bea
\text{tr} W_{\mu\nu} W^{\mu\nu} &\supset&   \partial V_T V_T V_T  \, ,   \label{sm11}\\[.2cm]
(D_\mu H)^\dagger D^\mu H &\supset&    \partial V_{L} V_{T} V_{L} +    v V_{T} V_{T} V_{L} \label{sm12} \, ,
\eea
where we have neglected SM coupling constants as well as ${\cal O}(1)$ numerical  factors. In (\ref{sm11},\ref{sm12}) we have also suppressed the Lorentz index contractions and denoted by $V$  either the $W$ or $Z$ vector boson. A one line calculation shows that the above TGC lead to $s$-channel amplitudes with the  leading  energy growth
\be
{\cal  M}\left(q\bar q \rightarrow V_TW_T^+\right) \sim   E^0  \  , \ \
{\cal  M}\left(q\bar q \rightarrow V_LW_L^+\right) \sim   E^0  \ , \ \
{\cal  M}\left(q\bar q \rightarrow V_TW_L^+/ V_L W_T^+\right) \sim  \frac{ v}{E} \, , \label{smgrowth}
\ee
where $E$ is the center of mass energy of the diboson system. The same asymptotic behavior is found for $W^-Z$ final states. In \reef{smgrowth} we are working in the limit of  massless light quarks, so that these only couple to the transverse gauge bosons, and we neglected subleading $\log (E)$ terms from loop corrections.
The process $q\bar q \rightarrow V_TW_T$ is also mediated by $t$,$u$-channel diagrams that have  the same energy growth as the $s$-channel in \reef{smgrowth}.

Next we discuss the energy growth of tree-level amplitudes involving  one insertion of the anomalous TGCs $\{\delta g_{1,Z}, \delta \kappa_Z, \lambda_Z \}$, defined in (\ref{type1},\ref{type2}). For this purpose, it is convenient to parametrize them in terms of the following dimension six operators,
 \be
 O_{HB}=  i g^\prime (D^\mu H)^\dagger D^\nu H B_{\mu\nu}\, , \ O_{HW}=  i g (D^\mu H)^\dagger\sigma^a D^\nu H  W^a_{\mu\nu}  \, ,    \ O_{3W}=   \frac{g}{3!}  \eps_{abc}  W_{\mu}^{a \, \nu}W_{\nu}^{b\, \rho}W_{\rho}^{c\, \mu} \ ,
 \label{silh1}
 \ee
which map onto the triplet  $\{\delta g_{1,Z},\, \delta \kappa_Z\, , \lambda_Z\}$ as follows
\be
\lambda_Z =   \frac{m_W^2}{\Lambda^2} c_{3W} \ , \quad \delta g_{1,Z} =\frac{m_Z^2}{\Lambda^2} c_{HW}\ , \quad \delta \kappa_Z = \frac{m_W^2}{\Lambda^2}  \left(c_{HW}-\tan^2  \th c_{HB}  \right) \ . \quad\label{atgcsilh}
\ee
 In principle one could use other sets of operators to parametrize deviations in the physics of $q\bar q\rightarrow WW/WZ$ production. However, it is important to realize that after taking into account the constraints from LEP-1, the main possible deviations in diboson production are due to modifications on the SM triple gauge vertices \cite{Elias-Miro:2013mua,Pomarol:2013zra}.~\footnote{Note that the commonly used SILH basis, apart from the operators of 
\reef{silh1}, also includes a further  operator contributing to the aTGC: $O_W= D^\mu 
W_{\mu}^{\ \nu}H D_\nu H +h.c.$. For our purposes though, it is enough to  use  \reef{silh1}  in 
order to   capture the high energy behavior. Our results will be presented  in terms of 
$\{\delta g_{1,Z},\, \delta \kappa_Z\, , \lambda_Z\}$, which can be mapped into any  other basis.} 
See also \cite{Falkowski:2014tna} where this result is studied using different bases of dimension six operators.

The operators in \reef{silh1} include the following TGCs
\bea
O_{HB}&\supset&     \partial W_L \partial Z_T \partial W_L+    v W_T \partial Z_T \partial W_L +   v^2 W_T \partial Z_T W_T+\dots \label{ohb} \, , \\[.2cm]
O_{HW}&\supset&  \partial V_L \partial V_T \partial V_L+  v V_T \partial V_T \partial V_L +    v^2 V_T \partial V_T V_T +\dots   \label{ohw} \, , \\[.2cm]
O_{3W} &\supset&    \partial V_T \partial V_T \partial V_T + \dots   \, , \label{o3w}
\eea
where ellipses denote interactions that either involve a photon or are not of the triple gauge type. Note that in (\ref{ohb})-(\ref{o3w}) we have neglected SM couplings as well as numerical ${\cal O}(1)$ factors. 
At large energies the leading processes mediated by the interactions in  \reef{ohb}-\reef{o3w}  are
\bea
{\cal  M}\left(q\bar q \rightarrow W_L^-W_L^+\right)& \sim&   E^2/\Lambda^2\    c_{HB}+ E^2/\Lambda^2\  c_{HW}  \sim  E^2/m_W^2\ \delta g_{1,Z} +  E^2/m_W^2\  \delta \kappa_{Z} \  , \label{bsmgrowth1} \\
{\cal  M}\left(q\bar q \rightarrow Z_LW_L^+\right) &\sim & E^2/\Lambda^2 \   c_{HW}      =E^2/m_Z^2 \ \delta g_{1,Z}\  , \label{bsmgrowth2}\\
{\cal  M}\left(q\bar q \rightarrow V_TW_T^+\right) &\sim& E^2/\Lambda^2 \ c_{3W}   = E^2/m_W^2\ \lambda_Z \label{bsmgrowth3}     \, ,
\eea
where we used \reef{atgcsilh} and omitted constant factors in front of the TGCs.
The same leading energy growth is found  by replacing $W^-\leftrightarrow W^+$ in the final state of \reef{bsmgrowth2}.
 Interestingly, $\delta\kappa_Z/c_{HB}$ contributes  at the order of $E^2$ only to the process \reef{bsmgrowth1}.
 The leading contribution of $\delta\kappa_Z$ to  $q\bar q\to W Z$ appears for the polarizations ${\cal  M}\left(q\bar q \rightarrow Z_T W_L^+\right)$ and scales as $\sim v E/\Lambda^2$. 
This follows from the fact that at leading order in energy 
only the transverse polarization of the $Z$ boson enters in $O_{HB}$. 

Next we discuss the generic properties of the production cross sections in the presence of these BSM amplitudes.

\subsection{Accuracy obstruction}

In general, the $2\to 2$ scattering cross section in the presence of irrelevant operators scales as
\bea
\label{eq:sigtt}
\begin{split}
\sigma(q\bar q \to V V)\sim \frac{g_{\text{SM}}^4}{E^2}\bigg[ 
 1 & +\overbrace{c_i\frac{E^2}{\Lambda^2}}^\text{BSM$_6\times\,$SM}
       +\overbrace{c_i^2 \frac{E^4}{\Lambda^4}}^\text{BSM$_6$$^2$}+ \dots \bigg]\, , 
\end{split} \label{genxsec}
\eea
where the first factor $g_\text{SM}^4/E^2$ accounts for the energy flux of the initial quarks, and we have omitted numerical factors. In \reef{genxsec} we explicitly indicated dimension six squared  and SM-dimension six interference terms, and  ellipses stand for  higher order corrections from operators of dimensions $\geq 8$.~\footnote{Note that operators of dimension 7 necessarily violate either baryon or lepton number. We assume the scale of such symmetry violation to be very large and therefore irrelevant for diboson physics at the LHC.} 
However, the operator $O_{3W}$ (i.e. the $\lambda_Z$ deformation) is special because the interference between the  amplitude ${\cal  M}\left(q\bar q \rightarrow V_TW_T^+\right) \sim   E^0 $ in \reef{smgrowth}   and  ${\cal  M}\left(q\bar q \rightarrow V_TW_T^+\right) \sim c_{3W}  E^2$ in  \reef{bsmgrowth3}  is suppressed and the scaling of the $\text{BSM}_6\times\text{SM}$ piece is softer. This is a consequence of the  helicity selection rules    \cite{Azatov:2016sqh}  as we will now review.~\footnote{See \cite{Dixon:1993xd} for a pioneering discussion of this effect in the context of QCD.} 

 The non-interference of the diboson production amplitude through $O_{3W}$ and the SM can be understood by first taking the limit where the masses of the electroweak gauge bosons are zero, namely we focus on transverse polarizations only.  
In this limit the tree-level SM process $q\bar q\to VV$  is only non-zero if the transverse helicities of the vector boson are opposite $(\pm,\mp)$.~\footnote{More generally, this   follows from the Maximally Helicity Violation (MHV) helicity selection rules, see for instance \cite{Dixon:1996wi}.}
At the same time though, the operator $O_{3W}$ in \reef{silh1}  leads to a triple gauge vertex where all three gauge bosons have the same helicity. 
A quick way to check this  is to write the field strength in terms of spinor indices $W_{\mu\nu} \sigma^\mu_{\alpha\dot\alpha} \sigma^\nu_{\beta\dot\beta}= w_{\alpha\beta}\bar\eps_{\dot\alpha\dot\beta}+ \bar w_{\dot\alpha\dot\beta}\eps_{\alpha\beta}$, where as usual the tensors $\eps$ and $\bar\eps$ are used to raise $\alpha$ and $\dot\alpha$ indices, respectively. 
$O_{3W}$ in \reef{silh1} can be written terms of the $w/\bar w$ fields is given as
\be
O_{3W}  \propto w_\alpha^{\ \beta} w_\beta^{\ \gamma} w_\gamma^{\ \alpha} +\bar w_{\dot\alpha}^{\ \dot \beta}\bar w_{\dot \beta}^{\ \dot \gamma} \bar w_{\dot \gamma}^{\ \dot\alpha} \, . \label{o3w2}
\ee
Each antisymmetric tensor field $w/\bar w$ can create a massless particle of spin $+1/-1$, respectively, and therefore  diboson production through \reef{o3w2} leads to vector bosons with helicity $(\pm,\pm)$. Thus, at tree level we have that
\bea
q\bar q& \longrightarrow &V_{T_\pm}V_{T_\mp} \quad \text{(in the SM)} \, , \label{prod1}\\
q\bar q& \longrightarrow &V_{T_\pm}V_{T_\pm} \quad \text{(with $O_{3W}$ insertion)} \label{prod2} \, . 
\eea
Since the final diboson states in (\ref{prod1},\ref{prod2}) are different, there is no interference between both amplitudes. 
This statement is exactly true in the massless limit. However, two mass insertions $m_W\partial_\mu G^+W^{- \mu}$, $m_Z\partial_\mu G^0Z^{ \mu}$ can be used to flip the helicity of the final states, leading to a non-zero interference between (\ref{prod1},\ref{prod2}). Flipping the helicity costs a factor $m_W^2/E^2$. Then, the  leading cross section for diboson production  in the limit $E\gg m_W$ is given by,
\be
\sigma ( q\bar q \rightarrow V_{T}V_{T} ) \sim \frac{g_\text{SM}^4}{E^2}\bigg[  1 +  c_{3W}\frac{m_V^2}{\Lambda^2}  +c_{3W}^2 \frac{E^4}{\Lambda^4}   \bigg]  \, . \label{supp}
\ee
The important point to notice is that the  second term of \reef{supp} has a suppressed  energy scaling  with respect to the general expectation in \reef{genxsec}.

This behavior makes EFT consistent  measurements of the $c_{3W}$  difficult.  Indeed, at the level of the dimension six operators the signal from the $O_{3W}$ will be subdominant compared to the contributions of the other TGCs, which will require further disentanglement of the transverse and longitudinal final state polarizations. But even more, assuming an ideal separation of the longitudinal polarizations we need to remain in the EFT validity range, namely in the parameter space where the contributions from the dimension eight operators can be safely ignored. 
For   the process  $q\bar q\to V_TV_T$  the  dimension eight contribution to the  cross section can be schematically written as
\bea
\label{eq:sigtt}
\begin{split}
\Delta\sigma_{\text{dim}=8}(q\bar q \to V_T V_T)\sim \frac{g_{\text{SM}}^4}{E^2}\bigg[ 
    \overbrace{c_8\frac{E^4}{\Lambda^4}}^\text{BSM$_8\times\,$SM} 
       +\overbrace{c_8^2\frac{E^8}{\Lambda^8}}^\text{BSM$_8$$^2$}+ \dots \bigg]\, .
\end{split} \label{dim8xsec}
\eea
Note that the $\text{BSM}_8\times \text{SM}$ piece scales as the  $\text{BSM}_6^2$ contribution,  $E^4/\Lambda^4$. Where we have  assumed that there is a interference between the SM and the new physics contributions at the level of the dimension eight operators. For the process $q\bar q\to V_T V_T$ this is indeed the case, consider for instance
  \be
 g D^{\nu}W^{\sigma\tau}W_{\nu\tau}D^{\mu}W_{\mu\sigma} \sim D^{\dot{\alpha}\alpha}\omega_{\alpha\beta}\bar{\omega}_{\dot{\alpha}\dot{\gamma}}D^{\dot{\gamma}\sigma}\omega_{\sigma}^{\beta} - D^{\alpha}_{\dot{\gamma}}\bar{\omega}^{\dot{\beta}\dot{\gamma}}\omega_{\alpha\gamma}D^{\sigma}_{\dot{\beta}}\omega_{\sigma}^{\gamma} + D^{\alpha}_{\dot{\gamma}}\omega^{\beta\gamma}\omega_{\alpha\gamma}D^{\dot{\sigma}}_{\beta}\bar{\omega}_{\dot{\sigma}}^{\dot{\gamma}} + \dots \, , \label{tgc8}
 \ee
 where ellipses denote terms with helicity configurations other than $\sim \om \om \bar \om$;
 or the operator
 \be
g^2 \,  (\bar q \gamma^\rho q)W_{\rho\nu} D^\mu W_{\mu\nu }  \sim  
 q^\alpha  \bar q_ {\dot \beta}   w_{\alpha}^{\ \beta} D_\beta^{\ \dot \alpha} \bar w_{\dot \alpha }^{\ \dot \beta} +\dots \, , \label{op8}
 \ee 
 written in terms of spinor indices. The latter  operator is a contact interaction contributing to $q\bar q\rightarrow VZ$ while   \reef{tgc8} is a modification of the TGC -- of the second type according to the discussion around (\ref{type1}-\ref{type2}). Note that both of them lead to final state bosons of helicities $(\pm,\mp)$, like in the SM.  
 
Then the  truncation  at the dimension six level \reef{supp} is valid if only  
\footnote{We are assuming that contributions of  operators of dimension higher than eight are even smaller.
}
\bea
\label{eq:sup}
 \hbox{max}\left(c_{3W}\frac{m_V^2}{\Lambda^2}, c_{3W}^2\frac{E^4}{\Lambda^4}\right)> \hbox{max}\left(c_{8}\frac{E^4}{\Lambda^4},c_{8}^2\frac{E^8}{\Lambda^8}\right).
\eea

Suppose  we will be able get rid of the interference suppression, then
this condition is replaced by
\bea
\label{eq:ns}
 \hbox{max}\left(c_{3W}\frac{E^2}{\Lambda^2}, c_{3W}^2\frac{E^4}{\Lambda^4}\right)> \hbox{max}\left(c_{8}\frac{E^4}{\Lambda^4},c_{8}^2\frac{E^8}{\Lambda^8}\right),
\eea
which is  less restrictive if $ c_{3W} E^2/\Lambda^2<1$ (given that at LHC $E> m_V$).

Another advantage of having a large interference term is that it leads to the better measurement of the sign of the Wilson coefficient, otherwise very weakly constrained. 
The importance of the improvement in \reef{eq:ns} depends on the actual values of the Wilson coefficients or in other words on the  UV completions of the given EFT. To make this discussion more concrete we present a few examples in the next subsection.

 \subsection{Power-counting examples}
The strength of the Wilson couplings can be estimated by a given set of  power-counting rules characterizing a possible UV completion. 
Power-counting schemes are useful to incorporate particular biases towards the kind of BSM physics we would like to prove. This is a perfectly legitimate strategy and very much the point of using an Effective Field Theory approach, allowing to parametrize altogether broad classes of models. Particular examples are weakly coupled renormalizable UV completions, Minimal Flavor Violation (MHV) \cite{DAmbrosio:2002vsn}, the Strongly Interacting Light Higgs (SILH) \cite{Giudice:2007fh}, flavor universal BSM physics (see e.g. \cite{Barbieri:2004qk}), etc. 
The power-counting schemes commonly used are imposed through arguments based on the symmetries or dynamics of the Action, such that possible radiative corrections violating the assumed power-counting scheme are kept small or understood.

For example, we may  assume that the UV completion is a renormalizable and weakly coupled QFT. Then,  the power-counting consist in classifying those operators that are loop generated v.s. those that are generated at tree-level  \cite{Arzt:1994gp,Elias-Miro:2013mua}. The latter are expected to be bigger because the  former are suppressed by  $1/(16\pi^2)$ factors. 
Then, for example if we have heavy vector-like fermions, we expect
\be
\label{eq:renorm}
c_{3W}\sim {\cal O}(1) \times g^2/(4\pi)^2\ , \quad c_{\reef{tgc8}} \sim  {\cal O}(1) \times g^2/(4\pi)^2 \, ,
\ee
where $c_{\reef{tgc8}}$ refers to the Wilson coefficient of the  dimension eight operator in \reef{tgc8}; the contribution to $c_{\reef{op8}}$ has a stronger loop suppression. This setup is somewhat pessimistic since the extra loop suppression makes it hard to prove $c_{3W}$ with the LHC sensitivity. In any case,   improvement from  \reef{eq:sup} to    \reef{eq:ns}  is 
\bea
E^2 < \Lambda m_W \ \longrightarrow   \ E<\Lambda \ .  \label{imp1}
\eea
As an other power-counting instance, one may envision a  scheme where for each extra-field strength that we add to the dimension four SM Lagrangian we pay a factor $g_*\lesssim 4\pi$. With this power-counting we obtain
\be
c_{3W}\sim g_*/g \ , \quad c_{\reef{tgc8}} \sim g_*/g \, ,\quad c_{\reef{op8}} \sim g_* g/ (16\pi^2) \, ,
\ee
where the $1/g$ factor is due to the normalization of $O_{3W}$ in \reef{silh1}. This power counting, called \emph{pure Remedios}, was introduced in \cite{Liu:2016idz}.~\footnote{In a nutshell, the construction is based on the   following observation. Consider the SM effective Lagrangian ${\cal L}_{EFT}={\cal L}_\text{Higgs} +{\cal L}_\psi + \frac{\Lambda^4}{g_*^2} L(\hat F_{\mu\nu}/\Lambda^2, \partial_\mu/\Lambda)$,  where the gauge-field strengths $\hat F_{\mu\nu}$ are not canonically normalized and we view $L$ as a functional that we expand in inverse powers of $\Lambda$. Then, it is technically natural to set $g_*\gg g$ in  ${\cal L}_{EFT}$  because as $g\rightarrow 0$  the  $SU(2)_L$ gauge symmetry acting on ${\cal L}_\text{EFT}$ is deformed into $SU(2)_L^\text{global}\rtimes U(1)^3_\text{gauge}$  -- we refer to  \cite{Liu:2016idz} for details.} 
This power-counting is more optimistic regarding possible LHC signals since $g_*$ can be naturally large. 
However, in this scenario there is no     improvement from  \reef{eq:sup} to    \reef{eq:ns}, and in both cases we find 
\bea
E< \Lambda \, . 
\eea
Lastly we will discuss one scale one coupling power-counting \cite{Giudice:2007fh}, which predicts
\bea
 c_{3W}\sim c_{\reef{tgc8}}\lesssim \frac{g_*}{g},~~ c_{\reef{op8}}\lesssim \frac{g_*^2}{g^2} .
\eea
In this case the improvement from  \reef{eq:sup} to    \reef{eq:ns}  would be
\bea
E<\left(\frac{g\Lambda^2 m_W^2 }{g_*}\right)^{1/4}\longrightarrow E< \Lambda \sqrt\frac{g}{g_*} \, . 
\eea

To conclude this subsection we would like to remind the reader that  EFT validity discussion needs some assumptions on power-counting 
(see for a recent discussion  \cite{Contino:2016jqw}). 
In the rest of the paper though, we do not commit to any of the aforementioned power-counting rules. We only assume perturbative, but otherwise arbitrary, Wilson coefficients.

 \subsection{Numerical cross-check}
\begin{figure}[t]
\begin{center}
\includegraphics[scale=.45]{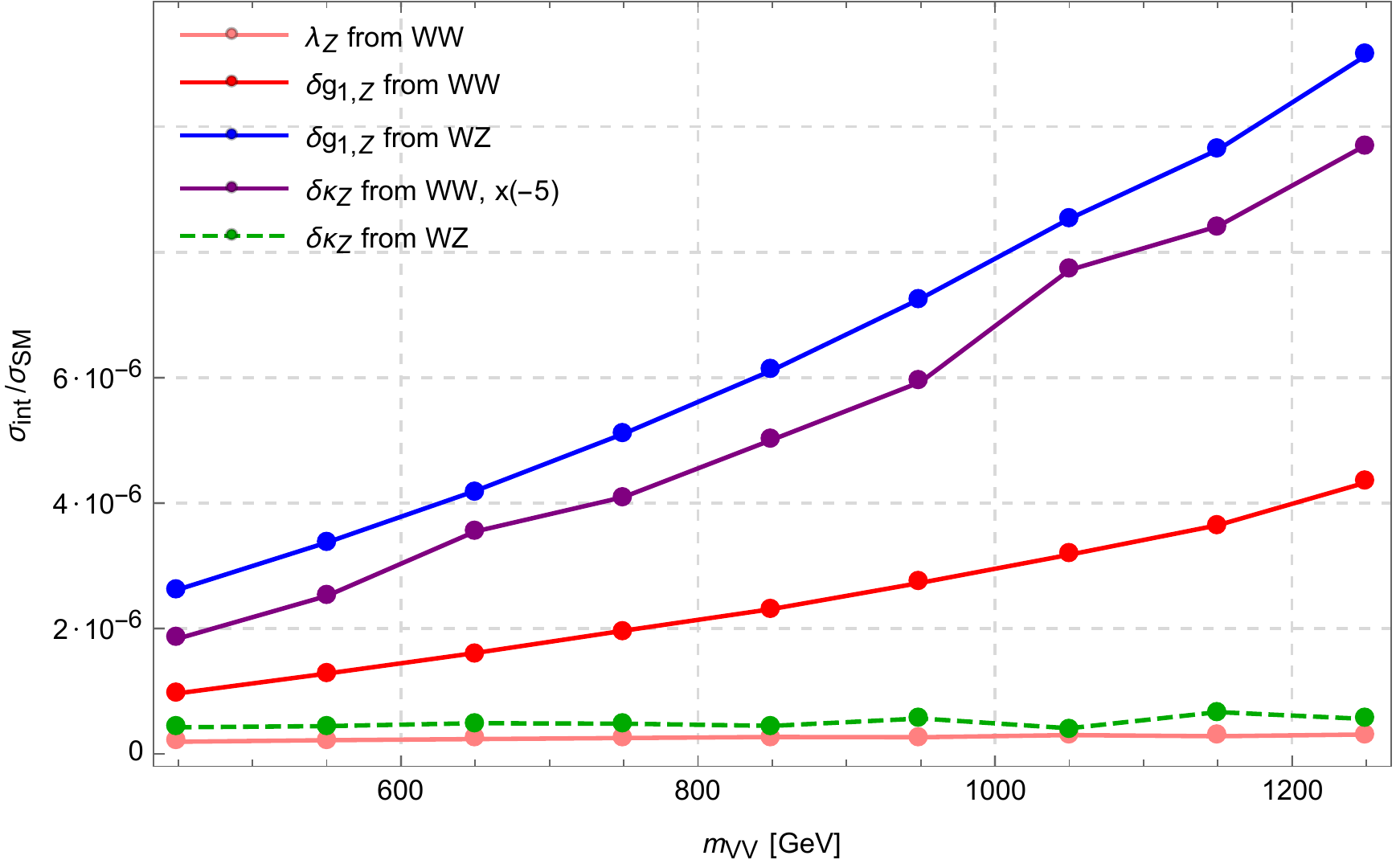} \quad
\includegraphics[scale=.465]{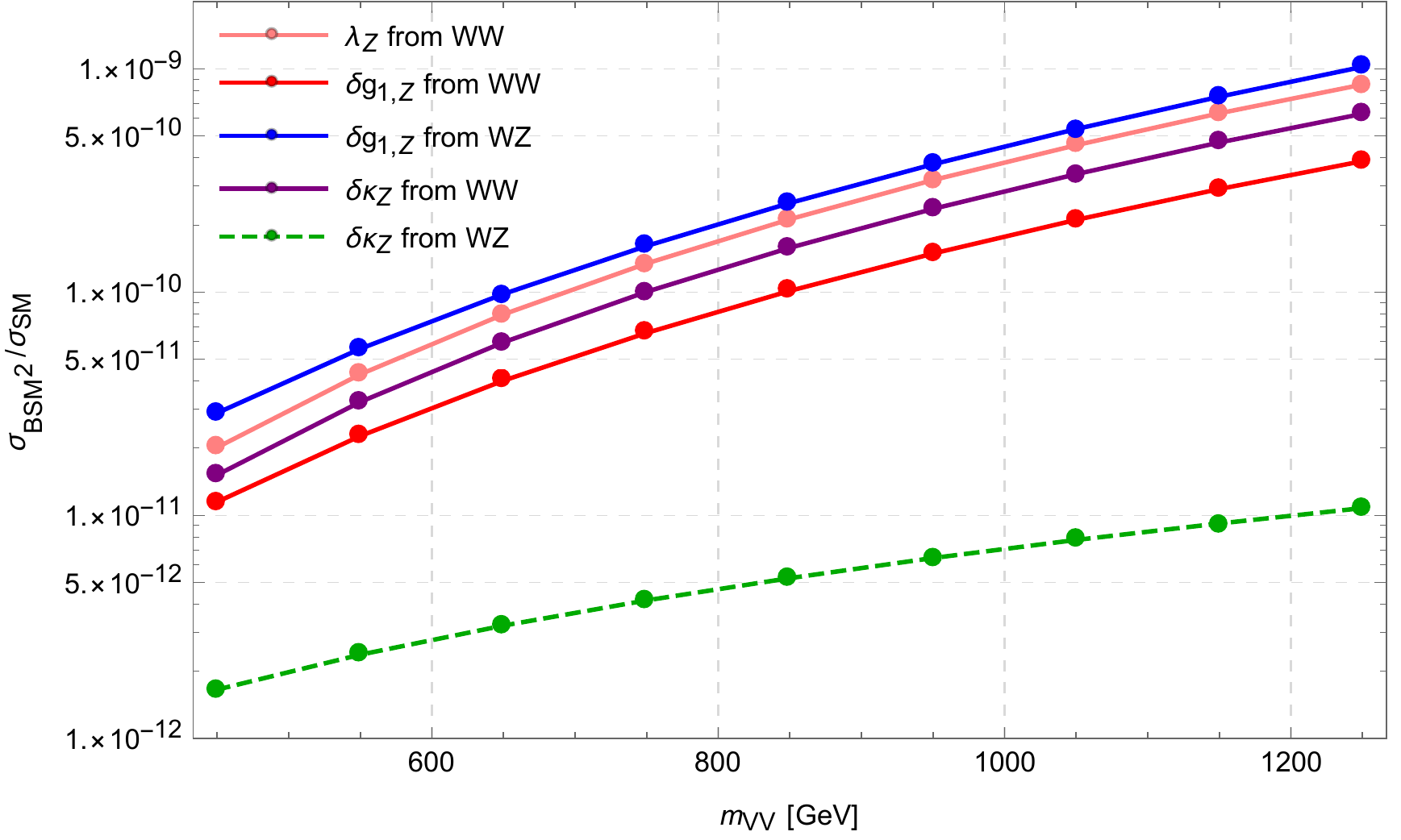}
\caption{Results from a \texttt{MadGraph5} simulation of the $pp\rightarrow VW$ process mediated by anomalous TGCs, see the main text.  The error bars of both plots due to statistical errors is within the width of the plotted lines.  We multiplied the line $\sigma_\text{int}/\sigma_\text{SM}$  of $\delta \kappa_Z$ from WW by $\times(-5)$ for illustrative reasons. 
\label{fig:edep}}
\end{center}
\end{figure}
In Fig. \ref{fig:edep}  we show the results of a \texttt{MadGraph5} \cite{Alwall:2014hca} simulation, using the \texttt{EWdim6} \cite{Degrande:2012wf} model~\footnote{Note that our definition in \reef{silh1} differs from the one of \cite{Degrande:2012wf}.} , for the process $pp\rightarrow VW$.  The parametric dependence of the cross section on the TGCs is given by
\be
\sigma_{q\bar q \rightarrow VW} = \sigma_\text{SM}+ \delta \, \sigma_\text{int} + \delta^2\, \sigma_{\text{BSM}^2} \ ,  \quad \text{with}\quad \delta = \{\delta g_{1,Z},\, \delta \kappa_Z , \,  \lambda_Z\} \, , \label{crossnum}
\ee
In Fig. \ref{fig:edep} we plot $\sigma_\text{int}/\sigma_\text{SM}$ (left) and $\sigma_{\text{BSM}^2}/\sigma_\text{SM}$ (right) for different anomalous TGCs as a function of the invariant mass $m_{VW}$ of the $VW$ final state system. 
Note that in this ratios the $g_\text{SM}^4/E^4$ factor in \reef{genxsec} cancels and we can read the scaling as a function of the energy from \reef{smgrowth} and (\ref{bsmgrowth1}-\ref{bsmgrowth3}).

The left plot of Fig.~\ref{fig:edep}  shows the energy scaling of $\sigma_\text{int}/\sigma_\text{SM}$. The red and purple lines confirm the quadratic growth expected from the $\delta g_{1,Z}$ and $\delta\kappa_Z$ contribution in \reef{bsmgrowth1}. 
The dashed green line shows no growth as a function of the energy, this confirms the discussion of (\ref{smgrowth},\ref{bsmgrowth2}). Namely, that for the final state $ZW$, 
the leading energy growth is only mediated by $\delta g_{1,Z}$ (blue line) but not by $\delta\kappa_Z$.
Lastly, on the same plot we show that $\sigma_\text{int}/\sigma_\text{SM}$ mediated by $\lambda_Z$ has no energy growth, confirming \reef{supp}. This later measurement comes from WW production, but a similar result for $\lambda_Z$ is obtained for WZ production. 

In Fig.~\ref{fig:edep} right, we show the energy dependence of $\sigma_{\text{BSM}^2}/\sigma_\text{SM}$, confirming the theoretical expectations. Namely, we find that for $VW$ production the factor $\sigma_{\text{BSM}^2}/\sigma_\text{SM}$ mediated by  $\lambda_Z$ and $\delta g_{1,Z}$ scale with the same power $E^4$. Then, regarding $\delta\kappa_Z$ the amplitude grows as $E^2$ for $WZ$ production while it scales as $E^4$ for $W^+W^-$ production --  this is the expectation from the squared amplitude $|{\cal  M}\left(q\bar q \rightarrow Z_TW_L^+/Z_LW_T^+\right)|^2 \sim v^2 E^2  \delta \kappa_Z^2$, see text after \reef{bsmgrowth3}.

\section{Solutions to the non-interference obstruction}
\label{sols}

In the previous section we showed  that for the $2\to 2$ processes 
 the interference between $O_{3W}$ and the SM is suppressed. In this section we will present two ways to overcome this suppression. For simplicity reasons in the rest of the paper  we will consider the case  when  only  $\lambda_Z$ deformation is present and the other anomalous TGCs are set to zero.

\subsection{Angular distributions}
\label{angdist}

\begin{figure}
\begin{center}
\includegraphics[scale=0.5]{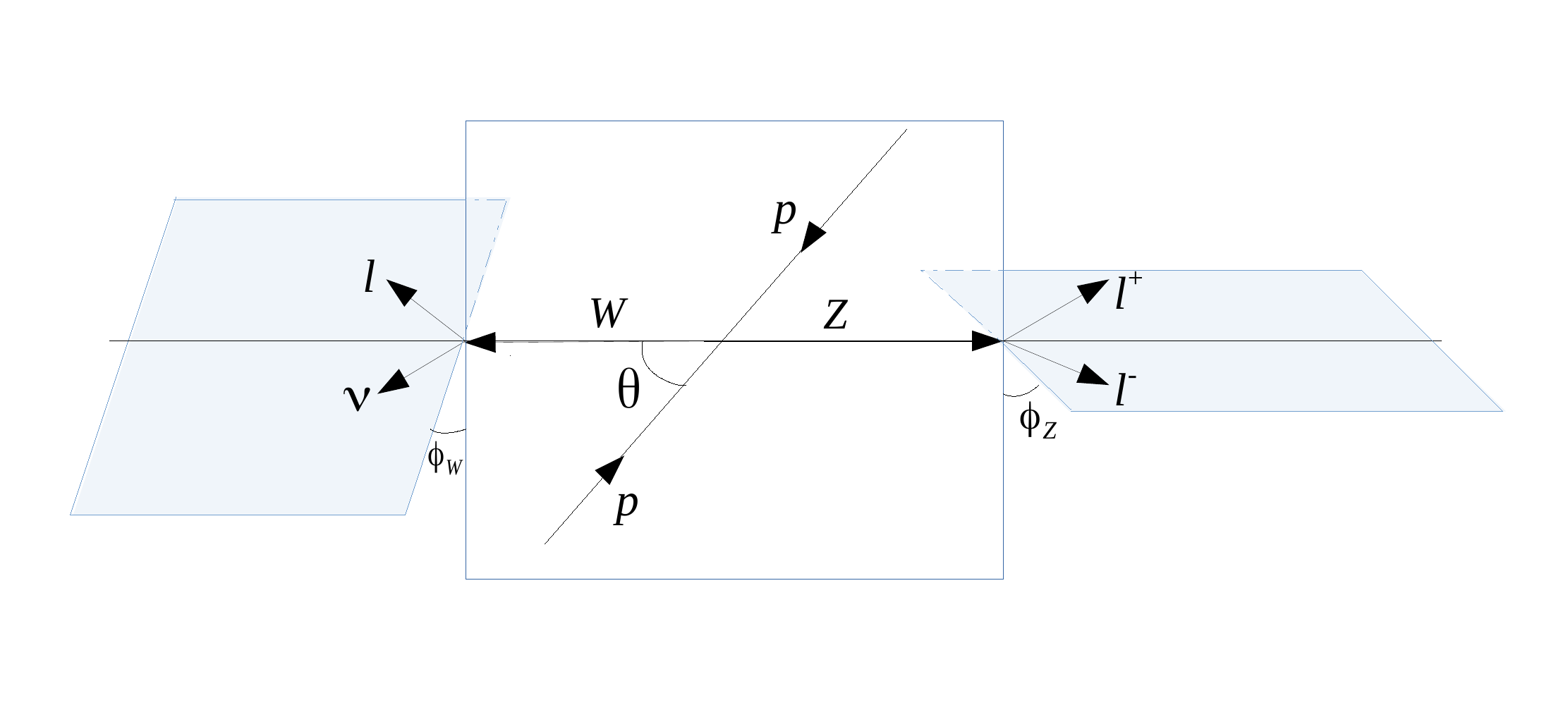}
\end{center}
\caption{Angles for $2\to 4$ scattering \label{fig:ang}}
\end{figure}

The first way of enhancing the interference term is by noting that in reality we are not looking at the $2\to 2$ process but at $2\to 4$, i.e. vector bosons decay into fermions $q\bar q\to VW\to 4 \psi$.  
Let us consider the differential cross section for the production of the polarized particles $W_{T+} l_- \bar l_+$ \footnote{ Similar ideas where proposed recently for the $W\gamma$ final state \cite{talksPPR}.}
\be
\frac{d\sigma(q\bar q \rightarrow W_{T_+} l_- \bar l_+)}{d\text{LIPS} } =
 \frac{1}{2 s}  \frac{ \left|\sum_i  
(  {\cal M}^\text{SM}_{q\bar q \rightarrow W_{T_+}Z_{i}}+
 {\cal M}^\text{BSM}_{q\bar q \rightarrow W_{T_+}Z_{i}}
)
  {\cal M}_{Z_{i}\to  l_- \bar l_+}   \right|^2}{(k_Z^2-m_Z^2)^2+m_Z^2\Gamma_Z^2} 
\,  , \label{xsecan}
\ee
where   sum runs over intermediate Z polarizations and $d\text{LIPS}\equiv (2\pi)^4\delta^4(\sum p_i -p_f) \prod_i {d^3 p_i}/\left(2 E_i(2\pi)^3\right)$ is the Lorentz Invariant differential 
Phase Space. We have factored out a Z-boson propagator, inputing the fact that all $Z$ 
polarizations  have the same mass and width. It is well known that at LHC SM process is 
dominated by the transverse polarizations \cite{Baur:1994ia}, so  for simplicity let us  ignore the 
contributions from the intermediate longitudinal $Z_L$ bosons.
Then in the narrow width approximation the leading contribution to the interference, i.e. the cross term $\text{SM}\times\text{BSM}$ in \reef{xsecan}  is given by:
\be
 \frac{\pi}{2 s}  \frac{\delta (s-m_Z^2)}{\Gamma_Z m_Z}   
  {\cal M}^\text{SM}_{q\bar q \rightarrow W_{T_+}Z_{T_-}} \left(  {\cal M}^\text{BSM}_{q\bar q \rightarrow W_{T_+}Z_{T_+}}  \right)^* {\cal M}_{Z_{T_-}\to  l_- \bar l_+} 
 {\cal M}_{Z_{T_+}\to   l_- \bar l_+}^*
    + h.c.  \, .  \label{dxsc1}
\ee
The interference cross section in \reef{dxsc1} scales with the function  ${\cal M}_{Z_{T_-}\to l_-\bar l_+} {\cal M}^*_{Z_{T_+}\to l_-\bar l_+}$. This in turn is modulated by the azimuthal angle $\phi_Z$ between the plane defined by the $Z$ decay  leptons 
and the scattering plane (formed by collision axis and $Z(W)$ bosons), see Fig.~\ref{fig:ang}. It is straightforward to compute \reef{dxsc1}, leading to
\be
 \frac{d\sigma_\text{int}(q\bar q \rightarrow W_{+} l_- \bar l_+)}{d\phi_Z}  \propto \cos(2\phi_Z)  \, . \label{ang1}
\ee

The derivation of  \reef{ang1}  is analogous if we consider the decay of the W gauge boson. Therefore,  the differential interference term   for the process $q\bar q\to VW\to 4 \psi$ is  unsuppressed and modulated as
\bea
 \frac{d\sigma_\text{int}(q\bar q \rightarrow W Z\rightarrow  4\psi)}{d\phi_Z\, d\phi_W}  \propto  \cos(2\phi_Z)+\cos(2\phi_W) , \label{simp}
\eea
where $\phi_{W,Z}$ are the corresponding azimuthal angles. 
Eqs.(\ref{ang1},\ref{simp}) are one of our main results. Namely, we would like to take advantage of the modulation of the interference term to prove the anomalous triple gauge coupling $\lambda_Z$. 
Similarly there is an effect of  interference  between the intermediate  longitudinal and the transverse vector bosons \cite{talksPPR}. The form of the  modulation is different from  \reef{simp}  and is proportional  $\propto \cos\phi_{W}\cos \phi_Z$,  
 however this effect is   suppressed due to the small SM cross section for the longitudinal vector bosons.

\begin{figure}
\begin{center}
\includegraphics[scale=0.646]{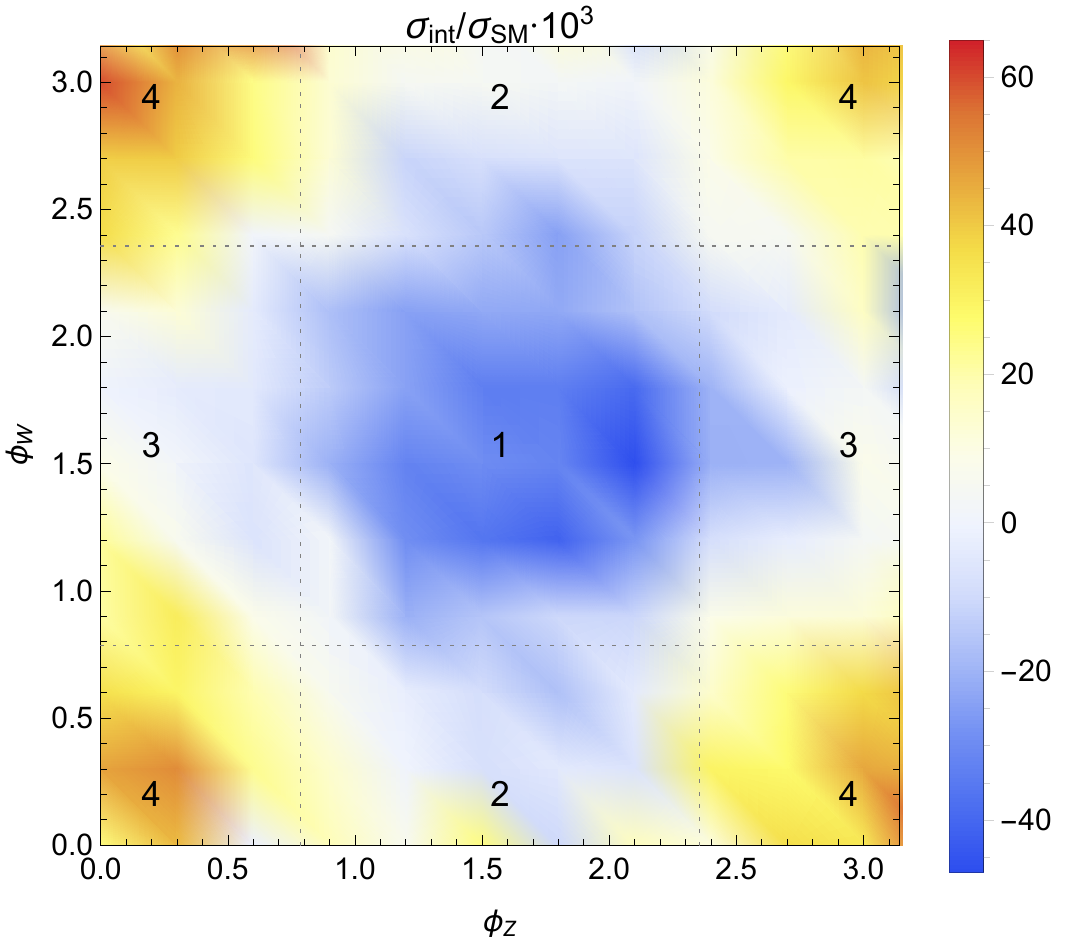} \quad 
\includegraphics[scale=0.53]{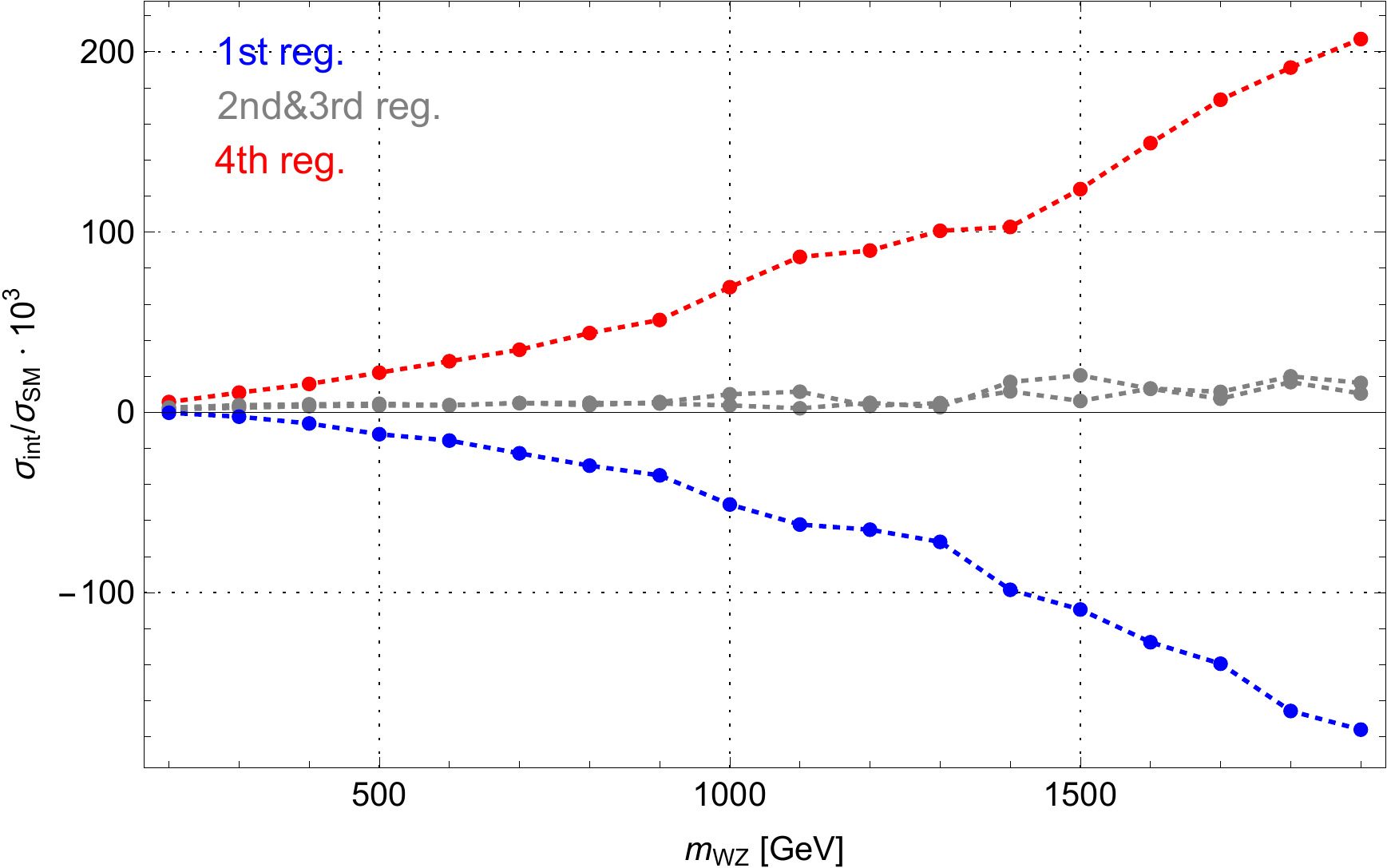}
\caption{\textbf{Left:} Differential interference cross section over SM one as a function the azimuthal angles $\phi_{W,Z}$ for the events with  $W-Z$ invariant mass $m_{WZ}\in[700,800]GeV$. \textbf{Right:} same quantity as a function of the $m_{WZ}$ binned according in the four bins defined in the left plot.
 \label{fig:angtheory}}
\end{center}
\end{figure}

Note that, naively, if the vector bosons are produced on-shell one would expect that vector bosons with different helicity contributions  should not interfere (or be suppressed by their width) even if we look at the decay products. 
Namely, one may expect that the interference  is further suppressed than if the same  $2\to 4$ amplitude was mediated by a $2 \to 2$ sub-process $q \bar q\rightarrow VW$ that does lead to a cross section containing an interference term. 
 However, this is not true, due to the basic fact that the both helicities  have the poles of the propagators at exactly the same energies.
Note that in the hypothetical case where  the $2\to 2$ process $  {\cal M}^\text{BSM}_{q\bar q \rightarrow W_+Z_-} \sim E^2/\Lambda^2$ was not suppressed, we would had gotten an analogous $\Gamma_Z/m_Z\rightarrow 0$ limit in \reef{dxsc1} where the amplitude would be instead controlled by the azimuthal angle of the function ${\cal M}_{Z_{T_-}\to l_-\bar l_+} {\cal M}^*_{Z_{T_-}\to l_-\bar l_+}$ (no modulation in $\phi_i$ in this case), but otherwise the energy growth would be the same.

We have performed a \texttt{MadGraph5} numerical simulation to test our theoretical expectations.   The results
 shown in Fig.~\ref{fig:angtheory}.
In the left plot we show the interference differential cross section over the SM cross section as a function of $\phi_Z$ and $\phi_W$. 
The shape of the function is as predicted by \reef{simp}. This suggests that we should bin the events into four categories depending on whether $\phi_i\in[\pi/4,3\pi/4]$. The results are shown on the right plot of Fig. \ref{fig:angtheory}. The upper red line and the lower blue line correspond to the 
categories with $\phi_{W,Z}\in[0,\pi/4]\cup [3\pi/4,\pi]$  and $\phi_{W,Z}\in[\pi/4,3\pi/4]$. We can see that there is a strong cancellation between these two contributions, however individually both of them grow with energy. So binning in azimuthal angles will increase
dramatically the sensitivity to the interference.

\subsection{Going beyond LO}

The non-interference of $\text{SM}\times\text{BSM}$ in diboson production through $\lambda_Z$ in the $2\to 2$ process applies at tree-level only. 
Higher order corrections, either in the form of loops or radiation, overcome the interference suppression and lead to a $\text{SM}\times\text{BSM}$ cross section piece that does grow with energy. This was first noticed in the context of QCD for the gluon operator $\sim G_{\mu}^{\ \nu}G_{\nu}^{\ \rho}G_{\rho}^{\ \mu}$  \cite{Dixon:1993xd}.
Here we apply this idea to the electroweak sector. 
The corrections from the virtual gluon will introduce the BSM-SM interference, however this 
effect will be suppressed by $\sim \frac{\alpha_s}{4\pi}$ compared to the angular 
modulation discussed in the previous section. Another possibility is to consider $2\to 3$ 
processes, namely the  production of the pair of the electroweak bosons with a  hard QCD jet $VV+j$ .Then  using Eq. \reef{o3w2} 
the BSM amplitudes have following helicity configuration,
\begin{center}
\includegraphics[scale=0.6]{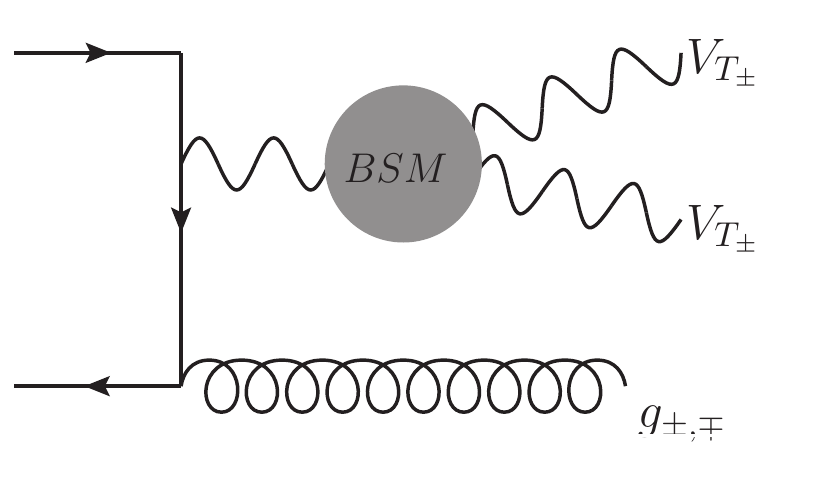} 
\end{center}
 where the gluon $g$ can take any polarization.
In the SM the same process has necessarily the helicity configuration 
\begin{center}
\includegraphics[scale=0.6]{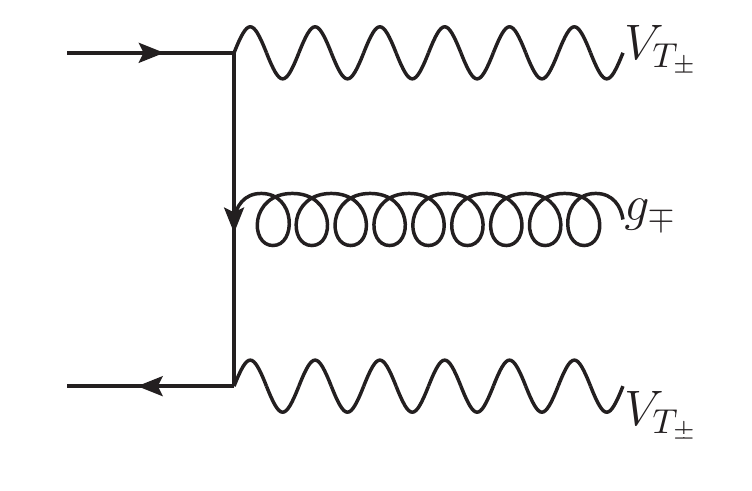} 
\end{center}
 i.e. it  can not be  of the Maximally Helicity Violating type. 
Thus, the  extra gluon radiation helps in sucking helicity allowing the same final state process as in $VV+j$ mediated by $O_{3W}$.
We find this simple observation interesting, since the requirement of extra radiation qualitatively changes the cross section behavior and provides a better handle on the interference terms.
Note also that the solution we are advocating in this section is complementary to the analysis presented in the section \ref{angdist}, in addition to the binning in the azimuthal angle we just require an extra hard jet.

Remember  that the interference effect becomes small both in the soft and collinear jet limits~\cite{Dixon:1993xd}. 
This is expected since interfering  SM amplitudes  $A(q\bar q\to V_{T_\pm} V_{T_\pm} g_\mp)$ cannot be generated from $A_{SM}(q\bar q\to VV)$ by splitting quark(anti-quark) line into $q(\bar q)\to q (\bar q) g$. So there will be no usual soft and collinear singularities corresponding to the poles of the splitting functions, which we have checked by explicit calculation.
 Then  the interference term in these limits, even if growing with energy, will be completely buried inside  the SM contribution.

\begin{figure}[htb!]
 \centering
 \includegraphics[scale=1.2]{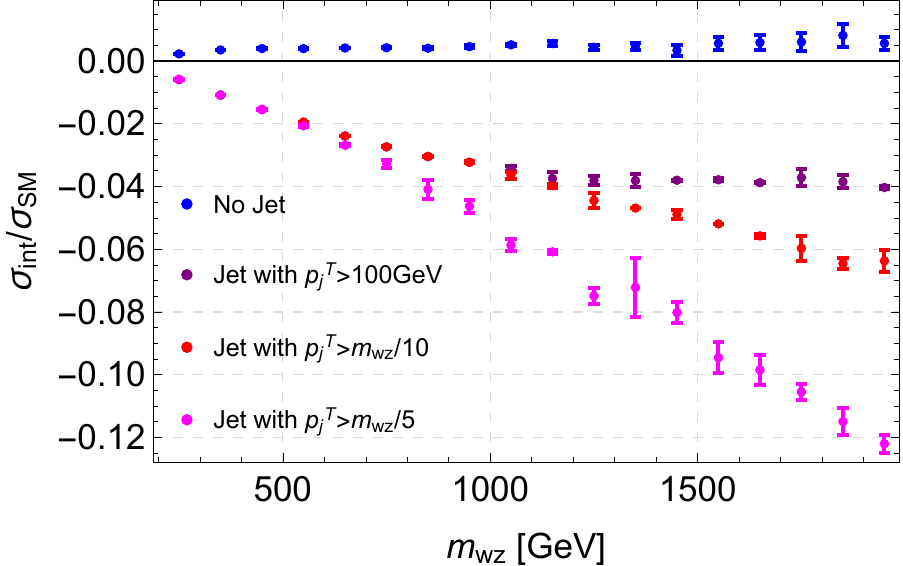}
\caption{
 \label{fig:a1a0mwz0jAndjDyn}
$\sigma_\text{int}/\sigma_\text{SM}$ as a function of $m_{WZ}$ for the process $pp\to WZ$ (blue) and the process $pp\to VW+j$, with $p^T_{j}> m_{WZ}/5$ (pink),  $p^T_{j}> m_{WZ}/10$ (red),  and  $p^T_{j}> 100~\text{GeV}$ (purple).}
\end{figure}

We cross-check the  theoretical expectations with a  \texttt{MadGraph5} simulation. 
In  Fig.~\ref{fig:a1a0mwz0jAndjDyn} we plot the ratio $\sigma_\text{int}/\sigma_\text{SM}$  for diboson production as a function of the invariant mass $m_{WZ}$, making various requirements on the extra gluon.
In blue we ask for no extra radiation which corresponds to the non-interference effect discussed in Fig.~\ref{fig:edep}.
In red and pink we require a hard gluon which takes a significant fraction of the diboson phase-space, $m_{WZ}/10$ and $m_{WZ}/5$ respectively. 
Importantly, the simulation shows the expected energy growth of the interference term. 
On the other hand, the purple curve does not show a steady growth of the energy. This is also expected since that curve is obtained by imposing a fixed lower cut on the jet 
$p^T$. As the energy of the diboson is increased  the extra jet
 becomes 
relatively soft and the energy growth is lost. 
We find by numerical simulations (see Fig. \ref{fig:a1a0mwz0jAndjDyn}) that we need to require something like $p^T_j\gtrsim \frac{m_{WZ}}{5}$ to have a quadratic growth with energy.
Error bars are due to the statistical treatment of the Monte Carlo (MC) simulation -- we regard them as small enough to convey our point.

\section{EFT validity}
\label{sec:EFTvalid}

\begin{figure}[t]
\begin{center}
\includegraphics[scale=0.7]{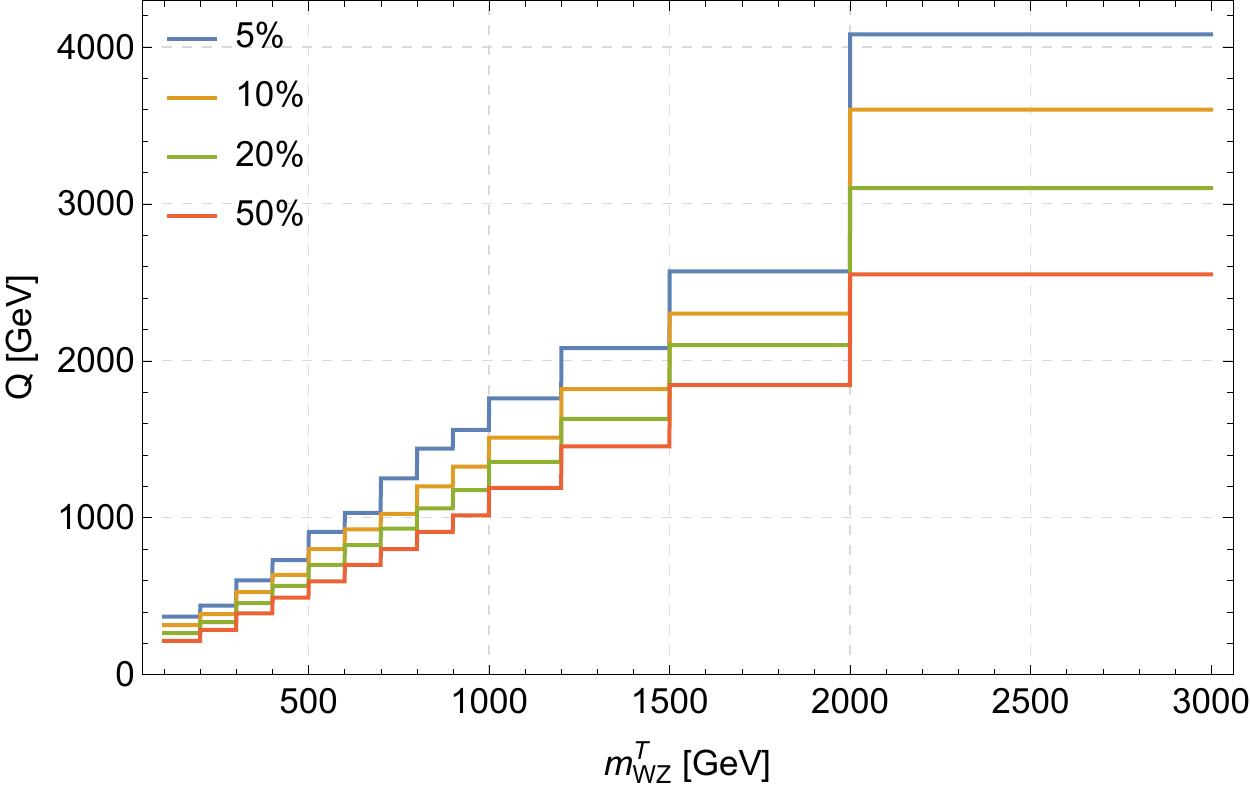}
\end{center}
\caption{We show, for the process $q\bar q\rightarrow WZ$ with $\lambda_Z$ turned on,  the leakage as a function of $m_{WZ}^T$, see main text for the definition.} \label{leakage}
\end{figure}

So far we were presenting the observables particularly sensitive to the SM$\times$BSM
interference term.
However this is not enough to ensure the validity of the EFT interpretation of diboson production at the LHC.
The convergence  of the EFT expansion  is controlled by the ratio of the  invariant mass 
of the diboson system over the new physics scale and thus $m_{VW}/\Lambda\ll 1$ 
should be satisfied.
However  at the LHC  it is hard to keep $m_{VW}/\Lambda$ fixed.
First, the precise  collision energy is unknown and not fixed, leading to an imprecise knowledge of $m_{VW}$ from event to event. 
Secondly and more importantly,  in many instances experimentalists  only reconstruct the visible decay products. Namely,  the $W-Z$ transverse mass
\be
m^T_{WZ}=\sqrt{(E_T^W+E_T^Z)^2-(p_x^W+p_x^Z)^2-(p_y^W+p_y^Z)^2}  \, , 
\ee
in the $WZ$ production
or the (visible) dilepton invariant mass
\be
m_{ll} = \sqrt{(p_{l^-}+p_{l^+})^2}  \, , 
\ee
of the $WW$ decay products. 
The invariant mass $m_{VW}$ of the diboson system is always greater or equal the visible invariant masses
$
m_{VW}\geq m_{ll}, \, m_{WZ}^T 
$. 
This implies that binning and cutting the distributions in terms of $m_{ll}/m_{WZ}^T$ variables does not allow to ensure $m_{VW}/\Lambda \ll 1$.
As an illustration of this point, in Fig.~\ref{leakage} we show the \emph{leakage}. This is defined as the percentage of the number of events  in a given $m_{WZ}^T$ (or $m_{ll}$) bin with invariant mass $m_{VW}$  larger than a certain scale $Q$. In equations,
\be
\text{Leakage} = \frac{ N_i(m_{VW}>Q)}{N_i} \times 100 \, , \label{lkg}
\ee
where $N_i$ is the total number of events in the given $m_{WZ}^T$ (or $m_{ll}$) bin. 
For instance,  the red line in the bin  $m_{WZ}^T\in [1500,2000]~\text{GeV}$ is interpreted as follows. Of all the events in that bin, $50\%$ of them  have an invariant mass $m_{WZ}\gtrsim 1800~\text{GeV}$.  These numbers were calculated using only the $\sigma_{BSM^2}$ term of the cross section, see \reef{crossnum}, which 
is the term  giving the largest leakage.

Naively, we can use the information in  Fig.~\ref{leakage} to set consistent bounds on the EFT.
For example, if we require $
\Lambda=2~\text{TeV}$ and the precision of the measurement $\lesssim O(1)\times 5\%$ we should keep the transverse mass bins only up to 1.5 TeV. This would work under the assumption that the leakage calculated using the dimension six operator squared provides a conservative estimate compared to the full UV complete model, namely that we do not have a very
 large number of events for some value of invariant mass $M_*>2$ TeV.  This assumption is 
for example
spoiled in the 
  presence of the narrow Bright-Wigner resonances and  the calculation with dimension six operators underestimates the  cross section and leakage by the factor  of
\bea
\frac{\sigma^{full}}{\sigma^{d=6}}\sim \frac{\pi \Lambda^2 }{\Gamma^2}, \label{parax}
\eea
 which becomes very large for narrow resonances 
 ($\Lambda,\, \Gamma$ are the mass and the width of the resonance)
At the same time in the more strongly coupled theories \reef{parax} is  only  of order one $O(1)$. 
Thus, under the assumption $\sigma^{full}/\sigma^{d=6}\lesssim O(1)$,  we can use the Fig.~\ref{leakage}  to find the correspondence between the transverse and invariant mass cut-offs once the precision of the measurement is specified.

The leakage can be made arbitrarily small  by simply assuming  a large enough value of $\Lambda$ in the EFT interpretation.  Then there is obviously no danger of narrow Breit-Wigner peaks, since the new particles would be too heavy to be produced at LHC.
However, this is somewhat dissatisfying because then LHC sensitivities only allow to prove Wilson coefficients that are on the verge of non-perturbativity, in order to compensate the large value of $\Lambda$.
For instance in \cite{Sirunyan:2017bey} bounds on the TGCs Wilson coefficients are of order $c_i\lesssim[-2.5,2.5]$~\footnote{We have rescaled the bounds of \cite{Sirunyan:2017bey} to our normalization in \reef{silh1}.}, with the cut-off  $\Lambda_{}=1$TeV.
This is done by analyzing   the whole range of $m_{VW}^T\approx [50,650]$ GeV,
and thus we expect  large number of the  events to have invariant masses $m_{VW}\gtrsim 1~\text{TeV}$.
 Then for the proper EFT interpretation we should set $\Lambda \gtrsim 2~\text{TeV}$, thus implying that the bound gets loosened roughly as $c_i\lesssim[-2.5,2.5]\longrightarrow  c_i \lesssim  4\times [-2.5,2.5]$, which pushes the EFT even further on the verge of non-perturbativity.

Next we will discuss another possible approach to perform a consistent EFT analysis. 
It allows to lower the cut-off  $\Lambda$ and hence be sensitive to somewhat less exotic theories, at least when the statistics is enlarged in the upcoming future. 

\subsection{Dealing with the leakage of high invariant mass events}

The idea consists in comparing  the observed cross section with the new physics expectation only in the constrained phase space satisfying the EFT validity requirements. This approach was originally suggested for the Dark Matter searches at LHC   \cite{Racco:2015dxa} and later applied for the anomalous TGCs measurements  \cite{Falkowski:2016cxu}. Next we discuss our implementation of these ideas.

In the standard analysis, for every  bin say in $m_{WZ}^T\in[m_{1}^T,m_2^T]$,  one would compare  
the observed number of events $n_\text{obs}$ with the theory prediction $M_\text{th}$, which in our case  reads
\be
 \label{eq:ntot}
M_\text{th}=n_\text{SM}+n_1 c_{3W}+n_{\text{BSM}^2} c_{3W}^2 \, , 
\ee
where $n_\text{SM}$ is the SM prediction, and $n_1$, $n_{\text{BSM}^2}$ come from the $\sigma_\text{int}$ and $\sigma_{\text{BSM}^2}$ pieces in \reef{crossnum}. 
In practice  this comparison can be done by evaluating the likelihood on a given bin by a Poisson distribution 
$ 
p(n_\text{obs}|M_\text{th}) = \frac{1}{n_\text{obs}!
}e^{-M_\text{th}} M_\text{th}^{n_\text{obs}}$. Note however that if we took this procedure we would be comparing $M_\text{th}$ with $n_\text{obs}$ for events were the formula $M_\text{th}$ is not valid unless the new physics scale $\Lambda$ is very large -- see the discussion of Fig.~\ref{leakage}.

Instead, what we will do is to compare the observed number of events  with the quantity $N_\text{th}$, which we define as follows:
\bea
 N_\text{th}=\left\{ 
\begin{array}{c}
\widetilde N_\text{th}    \quad \text{if}  \quad \widetilde N_\text{th} >a_0\\
n_\text{SM}   \quad \text{otherwise}  \quad 
\end{array}\right. \ , \quad 
 \label{eq:ncut}
\eea
where we define $\widetilde N_\text{th}= \tilde n_\text{SM}+\tilde n_1 c_{3W}+\tilde n_{\text{BSM}^2} c_{3W}^2$
with  $\tilde n_i$ is defined as $n_{i}|_{m_{inv}<\Lambda_\text{MC}}$, i.e. we restrict the expected number of events in the EFT to have invariant mass $m_{WZ}$ (or $m_{WW}$) below certain fixed cut-off scale $\Lambda_{\text{MC}}$.~\footnote{We are distinguishing the assumed cut-off scale $\Lambda_\text{MC}$ set in the MC simulation from the true value of $\Lambda$ in the SM EFT, which is of course an unknown constant of nature. Also note that $\Lambda_\text{MC}$ is analog to the scale $Q$ introduced in \reef{lkg}.} Thus, in practice  the likelihood  is modeled by $ 
p(n_\text{obs}| N_\text{th}) = \frac{1}{n_\text{obs}!
}e^{-{N}_\text{th}} {N}_\text{th}^{n_\text{obs}}$. 

The key question is whether the bounds obtained using  \reef{eq:ncut} lead to more conservative estimates than the ones which could come from the knowledge of  full theory. 
The number of events in the full theory is 
\be
N_\text{full theory} = \widetilde N_\text{th}   + \left[N_\text{full theory} \right]_{m_{inv}>\Lambda_{MC}},
\label{eq:nfull}
\ee
where we approximated the theory below $\Lambda_{MC}$ by the EFT expansion. Note that both terms in \reef{eq:nfull} are positive. Then, the bounds from \reef{eq:ncut} are conservative  only if
\bea
|n_{SM}- N_\text{th}|\leq|n_{SM}- N_\text{full theory}| \, ,
\label{cond}
\eea
condition that is always fulfilled with our definition of  $N_\text{th}$ in \reef{eq:ncut}.

Finally, let us note that in \cite{Falkowski:2016cxu} the choice of the theory is $ N_\text{th}=  n_\text{SM}+\tilde n_1 c_{3W}+\tilde n_{\text{BSM}^2} c_{3W}^2$, instead of \reef{eq:ncut}.  This amounts to modifying the BSM amplitudes by the "form factor"
\bea 
{\cal M}_{BSM}\to {\cal M}_{BSM}\times \theta (\Lambda_\text{MC}-m_{inv}) \, , \label{form1}
\eea
where the $\theta(x)$ is the Heaviside step function or any close function like $(1+e^{\alpha\left[\Lambda_\text{MC}-m_{inv}\right]/m_{inv}})^{-1}$ with $\alpha \gg 1$~\footnote{Note though that such function is not analytic in $\Lambda^{-1}_\text{MC}$.}. 
Then, equation \reef{cond} is fulfilled only if one assumes that  the deviations from the SM below and above $\Lambda_{MC}$ are of the same sign,  $\hbox{sign} (\Delta\sigma_{\hbox{BSM}} )|_{m_{inv}>\Lambda_{MC}}=\hbox{sign}(\Delta\sigma_{\hbox{BSM}})|_{m_{inv}<\Lambda_{MC}}$. Or in terms of the variables in \reef{eq:ncut}
\bea
\hbox{sign}(N_\text{full theory}-  n_\text{SM}-\tilde n_1 c_{3W}-\tilde n_{\text{BSM}^2} c_{3W}^2)&=&\hbox{sign}(\tilde n_1 c_{3W}+\tilde n_{\text{BSM}^2} c_{3W}^2 ) \, . \label{conddd}
\eea
Note that this condition is trivially satisfied when $\text{BSM}^2$ dominates the cross section, however it is not true once interference term is of the same size   \cite{Falkowski:2016cxu}.

At last we would like to comment about the procedure in the experimental study \cite{Aad:2016ett}. There,  a  different form-factor for the new physics contribution 
is used 
\bea
{\cal M}_{BSM}\to {\cal M}_{BSM}\times \frac{1}{\left(1+\frac{m_{inv}^2}{\Lambda_{MC}^2}\right)^2} \ .  \label{expform}
\eea
 The different form factors  would lead to  identical results for $\Lambda_{MC}\gg m_{inv}$, but there will be order one differences for the events with invariant mass close to the cut-off $\Lambda_{MC}$.
Also, note that while the UV assumptions are very clear when using \reef{form1} they are somewhat more obscure in \reef{expform}. The reason being that the fall-off of the form factor in \reef{expform} is not steep enough and its validity requires some  discussion or assumptions on the leakage along the lines we did at around \reef{parax}.

\section{Details of the collider simulation and statistical procedure}
\label{details}
In this section we explain our procedure for estimating the improvements of the LHC sensitivity due to the differential distributions proposed in the section  \ref{sols}. We have decided to look at the  cleanest decay channel in the pair production of the vector bosons, namely the process $pp\to W^\pm Z\to lll\nu$. In our analysis we have followed the signal selection procedure presented in the experimental work \cite{Aad:2016ett}. For the signal simulation we have used  \texttt{MadGraph5} \cite{Alwall:2014hca}  with the model \texttt{EWdim6} \cite{Degrande:2012wf} at LO \footnote{ 
One can perform the complete NLO study of the anomalous TGC using the model \texttt{EWdim6NLO} by C. Degrande. In our study however we have decided to ignore the effects of the virtual gluon,  which we believe to be phenomenologically less important  (see discussion in section \ref{sols}.2). For other QCD advances in SM and BSM calculations of the weak boson pair production see \cite{Campanario:2013wta,Campanario:2015nha,Campanario:2012fk,Grazzini:2016ctr,Grazzini:2016swo,Dawson:2013lya}
}.
 
 We have checked that our partonic level simulation reproduces the acceptance at the particle level $A_{WZ}=0.39$, for  $8~\text{TeV}$  \cite{Aad:2016ett}; it is defined as the ratio of the fiducial to the total cross section
\be
\sigma_{W^\pm Z}^{tot}=\frac{\sigma^{fid}_{W^\pm Z\to l' \nu ll}}{B_W B_Z A_{WZ}} \, .  \label{xsectot}
\ee
The fiducial cross section is defined as
\be
\sigma^{fid}_{W^\pm Z\to l' \nu ll}=\frac{N_{data}-N_{bkg}}{{\cal L}\,  C_{WZ}}\times\left( 1-\frac{N_\tau}{N_{all}}\right)\ ,  \label{fid}
\ee
where the factor $C_{WZ}$ simulates the detector efficiency $
C_{WZ}= N_{events}^{particle}/N_{events}^{detector} \approx 0.6$ \cite{Aad:2016ett}, 
and we approximate it to be flavor universal. 
 In \reef{xsectot} $B_i$ denote the corresponding branching fractions; while the factor  $N_\tau/N_\text{total}$ in \reef{fid} is the contribution of the leptons from $\tau$ decays which 
  \cite{Aad:2016ett} estimated to be of $\sim 4 \%$ and thus we will ignore it. ${\cal L}$ is the integrated Luminosity, below we report results for ${\cal L}=300~\text{fb}^{-1}$ and $3~\text{ab}^{-1}$.

We bin all the events according to their transverse mass $m_{WZ}^T$, and transverse momentum of the jet $p_j^T$.  In particular  $p_j^T$ is binned as
 \be
 p_j^T  =  [0,100],\, [100,300],\, [300,500],\, [500,\infty] \ \text{GeV} \, . 
 \ee
  For the events with $p_j^T<100$ GeV we also bin  the azimuthal angle $\phi_Z$ into two categories
  \be
   \phi_Z\in[\pi/4,3/4\pi] \quad \text{and} \quad \phi_Z\in[0,\pi/4]\cup[3\pi/4,\pi] \, . 
  \ee
The azimuthal angle $\phi_Z$ is defined here as an angle between
the plane spanned by $Z$ boson decay  leptons and the plane formed by the collission axis and the Z boson. 
   For the higher $p_j^T $ bins we have checked that the  binning in azimuthal angle results in little improvement of the bounds. The reason being that the modulation effect becomes sub-dominant compared to energy growth due to additional hard jet.

For each bin defined above we calculate the cross section in the presence of the $c_{3W}$ deformation according to the formulas (\ref{eq:ntot}-\ref{eq:ncut}) for three values of the invariant mass cut-off 
\be
\Lambda_\text{MC}= 1,\, 1.5,\, 2 \ \text{TeV} \ .
\ee 
These are reasonable choices in view of the current direct exclusion bounds. 

In order to reduce the fitting time we have used partonic level simulation to determine the coefficients in the (\ref{eq:ntot}-\ref{eq:ncut}).
For the bin $p_j^T\in[0,100]$ GeV we sum partonic level simulations with 0 jet and 1 jet with $ p_j^T\in[20,100]$ GeV.
We have checked that for the SM  input this approximation
agrees  well with the results obtained with \texttt{Madgraph/Pythia} \cite{Sjostrand:2007gs} interface with 
showering and jet matching. { One may worry  whether emission of a QCD jet can spoil the azimuthal angle modulation, however we have checked that even for relatively hard jets $p^T_j\lesssim 100$ GeV  angular modulation remains an important  effect. This makes our partonic simulation results  robust.}

{\renewcommand{\arraystretch}{1.15} \renewcommand\tabcolsep{6.7pt}
\begin{table}[t]
\begin{center}
\begin{tabular}{ L{2.5cm}  l  l l l l  C{2.5cm} }
\toprule
 & \multicolumn{2}{l}{Lumi.  $300~\text{fb}^{-1}$}&& \multicolumn{2}{l}{Lumi. $3000~\text{fb}^{-1}$} & \multirow{2}{*}{$Q~[\text{TeV}]$}\\
 \cmidrule{2-3} \cmidrule{5-6} 
    &$95\%~\text{CL}$ &$68\%~\text{CL}$ & &$95\%~\text{CL}$ &$68\%~\text{CL}$ & \\
\hline
 Excl. &[-1.06,1.11] & [-0.59,0.61] & &  [-0.44,0.45] & [-0.23,0.23] & \multirow{4}{*}{$1$}  \\[.0cm]
 Excl.,  linear & [-1.50,1.49] & [-0.76,0.76]&& [-0.48,0.48] & [-0.24,0.24] \\[.0cm]
 Incl. & [-1.29,1.27] & [-0.77,0.76]&&  [-0.69,0.67] & [-0.40,0.39] \\[.0cm]
 Incl., linear &[-4.27,4.27] & [-2.17,2.17]& & [-1.37,1.37] & [-0.70,0.70] \\[.0cm]
 \hline
 Excl.  &[-0.69,0.78] & [-0.39,0.45]  && [-0.31,0.35] & [-0.17,0.18]& \multirow{4}{*}{$1.5$}\\[.0cm]
 Excl., linear & [-1.22,1.19] & [-0.61,0.61]&& [-0.39,0.39] & [-0.20,0.20] \\[.0cm]
 Incl. &[-0.79,0.85] & [-0.46,0.52]&&  [-0.41,0.47] & [-0.24,0.29] \\[.0cm]
  Incl., linear  & [-3.97,3.92] & [-2.01,2.00]&  & [-1.27,1.26] & [-0.64,0.64]\\[.0cm]
 \hline
  Excl. &[-0.47,0.54] & [-0.27,0.31]&& [-0.22,0.26] & [-0.12,0.14]& \multirow{4}{*}{$2$} \\[.0cm]
 Excl., linear &[-1.03,0.99] & [-0.52,0.51] & & [-0.33,0.32] & [-0.17,0.17]\\[.0cm]
  Incl.  &  [-0.52,0.57] & [-0.30,0.34] && [-0.27,0.31] & [-0.15,0.19] \\[.0cm]
 Incl., linear & [-3.55,3.41] & [-1.79,1.75] &&  [-1.12,1.11] & [-0.57,0.57] \\[.0cm]
\bottomrule
\end{tabular}
\end{center}
\caption{Exclusive (Excl.) bounds on $c_{3W}/\Lambda^2\times\text{TeV}^2$ are obtain according to the method described in Sec.~\ref{details}, binning in $\phi_Z$ and $p_j^T$. Inclusive (Incl.): no binning and jet veto at $p_j^T\leq 100~\text{GeV}$. The bounds of the rows \emph{Excl./Incl., linear} are obtained by including only the linear terms in $c_{3W}$ BSM cross section. The total leakage in the various bins of $m_{WZ}^T$ is $\lesssim 5\%$ for each value of $Q$.}
\label{tab:bounds}
\end{table}
}

 For the backgrounds we have followed 
closely the results in \cite{Aad:2016ett}, where it was shown that the dominant 
background for the anomalous TGCs is the SM $W,Z$ boson production. The second 
most important background comes from the misidentified leptons $\sim 12\%$ and $ZZ$
 final state $\sim 7\%$ and  the contribution of the $t\bar t$ is at percent level. Since most 
of these backgrounds come from the $q\bar q$ initial state (except for $t\bar t$ which is small) at 14 TeV we expect a very similar situation. In our study we have decided to consider only the SM  weak boson production as a background, the other contributions will provide an additional increase of the background by $\sim 20\%$ and the  relaxations of the bounds by $\sim 10\%$, which we ignore in our study. For systematic uncertainties we use the results in \cite{Aad:2016ett}, where it was reported that the dominant errors come from the muon and electron identification efficiencies and it was estimated to be at the level of $2.4\%$. 
The statistical analysis is done using the Bayesian approach, where systematic errors are estimated using one  nuisance parameter $\xi$, normally  distributed
\bea
p( N_\text{th}| n_\text{obs}) \propto \int d \xi 
e^{-\xi N_\text{th}} \left(\xi N_\text{th}\right )^{n_\text{obs}}\exp \left[\frac{(\xi-1)^2}{2\sigma_{syst}^2}\right] \, . 
\eea

\section{Results}
\label{res}

We present our bounds on $c_{3W}/\Lambda^2$ in Table \ref{tab:bounds}. We report LHC prospects for $300~\text{fb}^{-1}$ as well as for $3~\text{ab}^{-1}$  luminosity (Lumi.) values.
Exclusive (Excl.) bounds are obtained according to the method described in Sec.~\ref{details}, binning in $\phi_Z$ and $p_j^T$, while inclusive (Incl.) corresponds to no binning in $\phi_Z$ and $p_j^T\leq 100~\text{GeV}$. 
 The total leakage in the various bins of $m_{WZ}^T$ is $\lesssim 5\%$ for each value of $Q$; such bins are selected using Fig.~\ref{leakage}.~\footnote{The scale $Q$ is roughly equal to the Monte-Carlo cut-off  $\Lambda_{MC}$, but see the discussion of  Fig.~\ref{leakage} and Table \ref{comps}.}

The bounds of the rows \emph{Excl./Incl., linear} are obtained by including only the linear terms in $c_{3W}$ in BSM piece of cross section.
 Generically, this later procedure is of course inconsistent. However,  comparing linear v.s. non-linear gives a sense of how much sensitive are the bounds to the quadratic piece term $\text{BSM}_6^2$ in the cross section \reef{genxsec}.
In this respect, note that  the exclusive analysis sensitivity to the linear terms has drastically increased compared to the inclusive one. 
For instance, the gain from the second to the first row is very mild, implying that the bound is mostly proving the interference term. 
Instead, the bounds from the third to the fourth row drastically decrease implying that the consistent bound of the third row is giving a lot of power to the quadratic pieces in $c_{3W}^2$.
This comparison illustrates the improvement from the differential distributions versus the inclusive analyses. 
Of course such a gain is always expected. However, in this case the improvement is dramatic because, as explained in section~\ref{sols}, the interference terms  of the  differential cross section have a qualitatively different behavior, namely they grow with the center of mass diboson energy.

\begin{figure}[h]
\begin{center}
\includegraphics[scale=0.55]{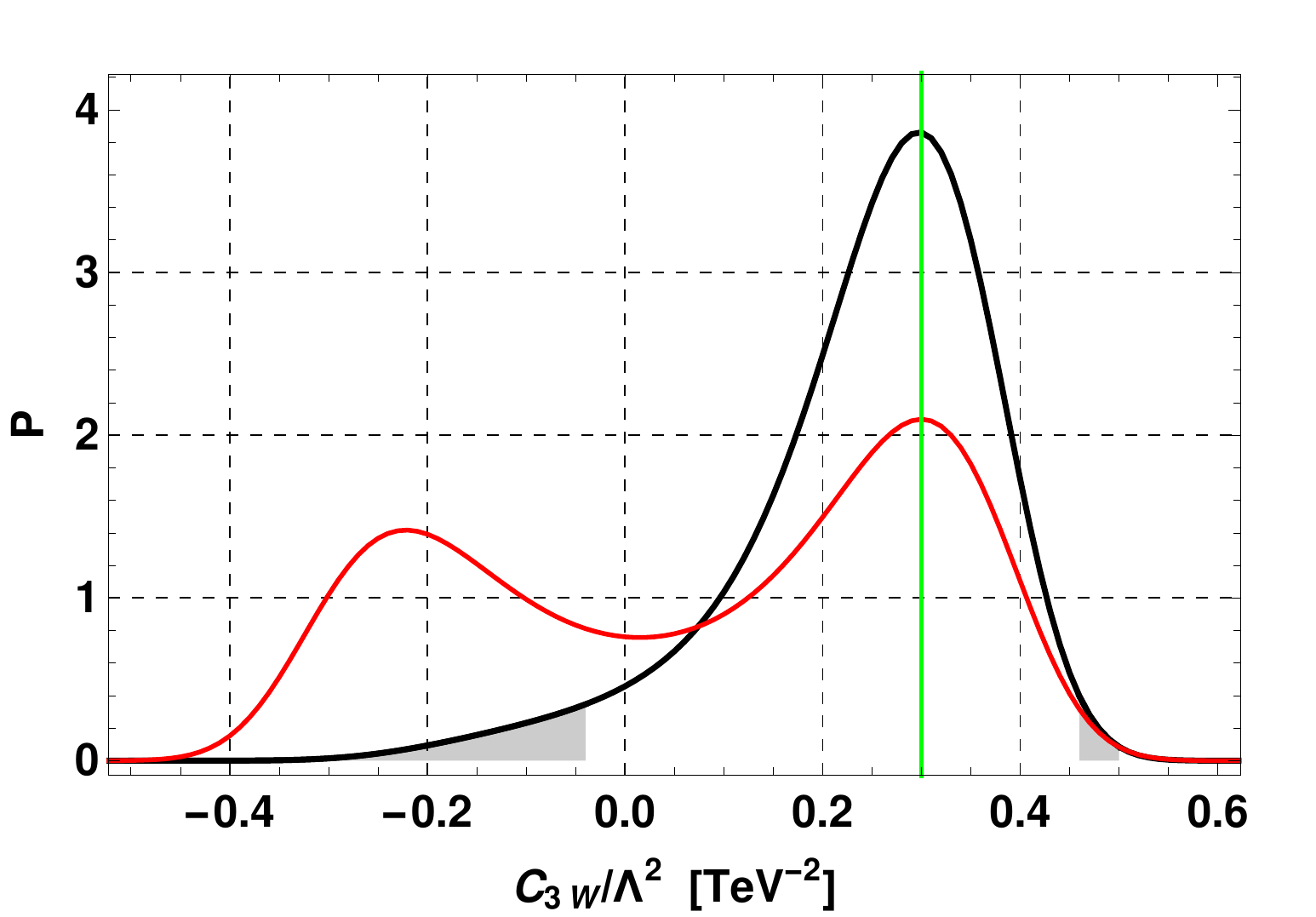}
\end{center}
\caption{Posterior probability for the inclusive and exclusive analysis after $3~\text{ab}^{-1}$ at LHC, see details in the main text.
\label{fig:sign}
}
\end{figure}

 This radical increase towards the sensitivity of the interference term is illustrated in Fig.~\ref{fig:sign}. There, we have injected a signal corresponding to the $c_{3W}/\Lambda^2=0.3~\text{TeV}^{-2}$. The red and black curves are posterior probabilities with $\Lambda_{MC}=2$~TeV and corresponding to inclusive and exclusive analysis
respectively
 (by inclusive we mean only binning in $m_{WZ}^T$ and ignoring high $p_j^T$ bins).
The curves are obtained by requiring the leakage to be $\lesssim 5\%$ as done in Table~\ref{tab:bounds}, (shaded grey area indicates the $95\%$ credibility intervals for the exclusive analysis).
We can clearly see that our variables will be able to access the  sign of the $c_{3W}$ 
Wilson coefficient otherwise hidden from the inclusive searches. 
Inspired by the  Fig.~\ref{fig:angtheory} we can see  that the following asymmetry variable   turns out to be very sensitive to the new physics contribution:
\bea
\label{eq:asym}
R_{\phi_Z}=\frac{N_{\phi_{Z}\in
[\pi/4,3\pi/4]}-N_{\phi_{Z}\in[0,\pi/4]\cup [3\pi/4,\pi]}}{N_{\phi_{Z}\in
[\pi/4,3\pi/4]}+N_{\phi_{Z}\in[0,\pi/4]\cup [3\pi/4,\pi]}}  \, . 
\eea
Indeed, we have checked that the SM contribution partially cancels, making $R_{\phi_Z}$ particularly sensitive to new physics contributions.

We would like to comment  for what  kind of theories our bounds are  
relevant. We can see that at most we are getting towards the constraint $c_{3W}/\Lambda^2\lesssim  0.2/ \text{TeV}^2$. Weakly coupled renormalizable theories  lead to the Wilson coefficients which are at least order 
of magnitude smaller \reef{eq:renorm}, unless we are dealing with abnormally  large multiplicities of new electroweak states just above the LHC reach. At the same time more strongly coupled theories can lead to the larger values of Wilson coefficients in the ball park of the LHC precision.

Table~\ref{tab:bounds} and Fig.~\ref{fig:sign} are our main final results.
We find that LHC at $3$ab$^{-1}$($300$fb$^{-1}$) will be able to constrain the $\lambda_Z$ aTGC coupling  to be 
\be
\lambda_Z \in[-0.0014,0.0016]~~ ([-0.0029,0.0034]) 
\ee
for the $95\%$ posterior probability interval for  $\Lambda_{MC}=2$ TeV.
Results for the other values of $\Lambda_{MC}$ can be trivially deduced from the Table~\ref{tab:bounds}).

 For the sake of completeness we also compare 
in Table~\ref{comps}  the  bounds on the Wilson coefficient obtained using the methods discussed in  the section \ref{sec:EFTvalid}. We can see that all methods lead to  results in the same ball park. 
Even though, the method of  \reef{eq:ncut} does not make any assumption on the nature of UV completion,   the sensitivity to the interference term is a bit worse than in the other two methods.

{\renewcommand{\arraystretch}{1.15} \renewcommand\tabcolsep{6.7pt}

\begin{table}
\begin{center}
\begin{tabular}{ l l  l l l  c}
\toprule
&  \multicolumn{2}{l}{Lumi. $300~\text{fb}^{-1}$}& \multicolumn{2}{l}{Lumi. $3000~\text{fb}^{-1}$}& \multirow{2}{*}{$Q~[\text{TeV}]$} \\
  & $95\%~\text{CL}$& $68\%~\text{CL}$  &  $95\%~\text{CL}$ &  $68\%~\text{CL}$ &    \\
\hline
Same as Tab.~\ref{tab:bounds}&[-1.06,1.11]&[-0.59,0.61]&[-0.44,0.45]&[-0.23,0.23] & \multirow{3}{*}{1}\\
 Use of \reef{eq:ncut} &[-1.59,1.55] &[-1.05,1.01]&[-1.17,1.06] &[-0.72,0.66]&\\
Method of \cite{Falkowski:2016cxu}&[-0.88,0.88]&[-0.50,0.50]&[-0.41,0.40] &[-0.22,0.22]&\\
\hline
Same as Tab.~\ref{tab:bounds} & [-0.69,0.78]&[-0.39,0.45]&[-0.31,0.35]&[-0.17,0.18]&   \multirow{3}{*}{1.5}\\
 Use of \reef{eq:ncut}&[-0.74,0.79] &[-0.48,0.50]&[-0.51,0.52]&[-0.34,0.30]\\
Method of \cite{Falkowski:2016cxu}&[-0.55,0.60]&[-0.32,0.35]&[-0.26,0.29]&[-0.15,0.16]
\\\hline
Same as Tab.~\ref{tab:bounds} &[-0.47,0.54]&[-0.27,0.31]&[-0.22,0.26]&[-0.12,0.14] &\multirow{3}{*}{2}\\
 Use of \reef{eq:ncut} &[-0.49,0.53] &[-0.30,0.34]&[-0.30,0.33]&[-0.20,0.20]\\
Method of \cite{Falkowski:2016cxu} &[-0.43,0.47]&[-0.24,0.27]&[-0.20,0.23]&[-0.12,0.13]\\
\bottomrule
 \end{tabular}
\end{center}
\caption{Comparison  of different methods.}\label{comps}
\end{table}
}


\section{Conclusions and outlook} 
\label{conclusion}
 We have discussed the prospects of the
measurements of the $c_{3W}$ Wilson coefficient ($\lambda_Z$ TGC) at LHC. This parameter was considered to be particularly difficult to test at hadron colliders due to the  suppressed interference effects. In our study we have shown that this suppression is not the case once the differential distributions are considered. 
In particular we have shown that this suppression can be overcome by studying the angular modulation in azimuthal angles  in \reef{simp}.
 Independently of this modulation  we have 
shown that requiring an  additional hard QCD jet leads to the energy growth of the interference between the SM and BSM contributions.

Looking at the cleanest $pp\to WZ\to lll\nu$ channel we have estimated the importance of these observables for the LHC  by calculating the prospects on the  
bounds at 300~fb$^{-1}$(3~ab$^{-1}$).
 Our simplified analysis by no means can be considered 
a complete experimental study, however the most important and robust results are the relative improvements of the measurements due to the angular modulations and the hard QCD jet distributions.
We have also discussed  the challenges of the consistent EFT analysis  for the TGC measurements at LHC.

The 
improvements in determination of $\lambda_Z$ due to the differential distributions turn out to be of the order of $15-25\%$ depending on the  assumptions on EFT cut-off.
 Even though this gain in precision   does not seem
 to be very big, the sensitivity to the interference term is significantly increased (factor of $\sim 3-4$), which makes the EFT expansion less model dependent as well as provides a handle on the sign of the Wilson coefficient. 
Of course it is not a novelty that the differential distributions improve the accuracy of the measurements. However in this case the improvement is particularly significant due to the  energy growth of the differential interference term.

In the future it would be interesting to use the differential distributions proposed 
 to perform a global EFT analysis  in order to find the best variables to distinguish between not only BSM and SM but also between different BSM contributions. 
Very similar azimuthal angle modulation will  appear every time there are amplitudes with different polarizations of the intermediate gauge bosons. These ideas will be explored in the future for the measurements of the other aTGCs.

It will be also interesting   to study the  azimuthal angle modulation for other $2\to 2$ processes that are  otherwise suppressed by the helicity selection rules,  like for example $V_TV_T\to V_{L,T}V_{L,T}$ .
 On the collider side, studies of the other decay channels as well as full inclusion of the NLO effects will be very important.

\section*{Acknowledgements} 

We would like to thank D. Barducci, G. Durieux, C. Grojean, M. Montull  for discussions. 


\appendix 

\section{Further details on the bounds}
In this appendix we compare the relative importance of the various differential observables on the constraints on $c_{3W}/\Lambda^2$. The results for $300(3000)~$fb$^{-1}$ are presented in the Table~\ref{tab:3000}. The labels {\it Excl./Incl. linear}  have exactly the same meaning as in the Table \ref{tab:bounds} . {\it No $\phi_Z$ binning} stands for binning only $p^T_j$ and {\it No $p_j^T$ binning} stands for using only the information in $p^T_j\in[0,100]$GeV category and the angular binning. We can see that both binning $p^T_j$ and $\phi_Z$ lead to the increase of sensitivity of the interference term with the later being  stronger. Table~\ref{tab:3000} is generated using the leakage $\lesssim 5\%$ for various $Q$ values. The procedure  of \cite{Falkowski:2016cxu} leads roughly to the same results and  the method of \reef{eq:ncut} shows lower sensitivity  on the interference term. Bin by bin information about the SM and BSM contributions can be available by request.

{\renewcommand\tabcolsep{6.7pt}  \renewcommand{\arraystretch}{.915} 
\begin{table}[h]
\begin{center}
\begin{tabular}{l l l l  l  c }
\toprule
& \multicolumn{2}{l}{Lumi. $300~\text{fb}^{-1}$} &  \multicolumn{2}{l}{Lumi. $3000~\text{fb}^{-1}$}  & \multirow{2}{*}{$Q~\text{[TeV]}$} \\
&$95\%~\text{CL}$ &$68\%~\text{CL}$  &$95\%~\text{CL}$ &$68\%~\text{CL}$ &\\
\hline
Excl. & [-1.06,1.11] & [-0.59,0.61]  & [-0.44,0.45] & [-0.23,0.23] & \multirow{8}{*}{$1$} \\
Excl., linear & [-1.50,1.49] & [-0.76,0.76]  & [-0.48,0.48] & [-0.24,0.24]  &\\
No $\phi_Z$ binning & [-1.19,1.20] & [-0.69,0.70]&  [-0.57,0.57] & [-0.32,0.31]& \\
No $\phi_Z$ binning, linear & [-2.28,2.22] & [-1.15,1.14] & [-0.74,0.73] & [-0.38,0.38] &\\
 No $p_j^T$ binning & [-1.14,1.17] & [-0.64,0.67]  & [-0.50,0.51] & [-0.27,0.27] &\\
No $p_j^T$ binning, linear & [-1.80,1.81] & [-0.91,0.92]  & [-0.57,0.57] & [-0.29,0.29] &\\
Incl. & [-1.29,1.27] & [-0.77,0.76]&  [-0.69,0.67] & [-0.40,0.39] &\\
Incl., linear & [-4.27,4.27] & [-2.17,2.17] & [-1.37,1.37] & [-0.70,0.70] & \\
 \hline
Excl. & [-0.69,0.78] & [-0.39,0.45] & [-0.31,0.35] & [-0.17,0.18] & \multirow{8}{*}{$1.5$}\\
Excl., linear  & [-1.22,1.19] & [-0.61,0.61]   & [-0.39,0.39] & [-0.20,0.20] \\
No $\phi_Z$ binning   & [-0.75,0.82] & [-0.43,0.49]   & [-0.37,0.43] & [-0.21,0.25] \\
No $\phi_Z$ binning, linear  & [-2.02,1.95] & [-1.02,1.00]   & [-0.65,0.64] & [-0.33,0.33] \\
 No $p_j^T$ binning   & [-0.73,0.80] & [-0.41,0.49]  & [-0.34,0.38] & [-0.19,0.20] \\
No $\phi_Z$ binning., linear   & [-1.43,1.40] & [-0.72,0.71]  & [-0.45,0.45] & [-0.23,0.23] \\
Incl.   & [-0.79,0.85] & [-0.46,0.52] & [-0.41,0.47] & [-0.24,0.29] \\
Incl., linear  & [-3.97,3.92] & [-2.01,2.00]    & [-1.27,1.26] & [-0.64,0.64] \\
 \hline
Excl. & [-0.47,0.54] & [-0.27,0.31] &  [-0.22,0.26] & [-0.12,0.14] & \multirow{8}{*}{$2$}\\
Excl., linear  & [-1.03,0.99] & [-0.52,0.51] &   [-0.33,0.32] & [-0.17,0.17] \\
No $\phi_Z$ binning  & [-0.50,0.56] & [-0.28,0.34]    & [-0.25,0.30] & [-0.14,0.18] \\
No $\phi_Z$ binning, linear & [-1.84,1.73] & [-0.92,0.89]  & [-0.59,0.58] & [-0.30,0.30] \\
 No $p_j^T$ binning  & [-0.49,0.55] & [-0.28,0.32]  & [-0.23,0.27] & [-0.13,0.15] \\
 No $p_j^T$ binning, linear  & [-1.18,1.12] & [-0.60,0.58]     & [-0.37,0.37] & [-0.19,0.19] \\
Incl.  & [-0.52,0.57] & [-0.30,0.34] &  [-0.27,0.31] & [-0.15,0.19] \\
Incl., linear   & [-3.55,3.41] & [-1.79,1.75]  & [-1.12,1.11] & [-0.57,0.57] \\
\bottomrule
\end{tabular}
\end{center}
\caption{Bounds on $c_{3W}/\Lambda^2$. The total leakage in the various bins of $m_{WZ}^T$ is $\lesssim5\%$.     \label{tab:3000} } 
\end{table}
}


 \small
 \bibliography{nlo-eft-Biblio}

\end{document}